\documentclass[showpacs,preprintnumbers]{aa}
\usepackage{natbib,twoopt}
\usepackage{graphicx,units,multirow,subfigure,color,amsmath,amssymb,epsfig,makecell}
\usepackage{courier}
\usepackage[varg]{txfonts}
\usepackage[usenames,dvipsnames,svgnames,table]{xcolor}
\usepackage[absolute,overlay]{textpos}

\usepackage[normalem]{ulem}

\bibpunct{(}{)}{;}{a}{}{,} 

\definecolor{light-gray}{gray}{0.95}
\definecolor{dark-gray}{gray}{0.4}

\usepackage[pdftex,             
            breaklinks=true,%
            colorlinks=true,%
            citecolor=blue,
            pdfauthor={Ferronato Bueno et al.},%
            pdftitle={Analysis of AMS-02 F/Si data}%
           ]{hyperref}

\newcommand{\BIG}{{\sf BIG}}
\newcommand{\SLIM}{{\sf SLIM}}
\newcommand{\QUAINT}{{\sf QUAINT}}
\newcommand{\yieldx}{{\sc yieldx}}
\newcommand{\wnew}{{\sc wnew}}
\newcommand{\exfor}{{\sc exfor}}
\newcommand{\galprop}{{\sc Galprop}}
\newcommand{\crdb}{{\sc crdb}}
\newcommand{\usine}{{\sc usine}}
\newcommand{\minuit}{{\sc minuit}}

\newcommand{\usinegen}{{\sc usine}}

\newcommand{\xsS}{{\tt S01}}
\newcommand{\xsWa}{{\tt W98}}
\newcommand{\xsW}{{\tt W03}}
\newcommand{\xsGalxii}{{\tt Galp--opt12}}
\newcommand{\xsGalxxii}{{\tt Galp--opt22}}
\newcommand{\optxii}{{\tt OPT12}}
\newcommand{\optxiiupxxii}{{\tt OPT12up22}}
\newcommand{\optxxii}{{\tt OPT22}}

\newcommand{\chidof}{\ensuremath{\chi^2/}\textrm{dof}}
\newcommand{\chimindof}{\ensuremath{\chi^2_{\rm min}}/{\rm dof}}

\newcommand{\chipernui}{\ensuremath{\chi^2_{\rm nui}/{n_{\rm nui}}}}
\newcommand{\veryshortarrow}[1][3pt]{\,\mathrel{%
   \vcenter{\hbox{\rule[-.5\fontdimen8\textfont3]{#1}{\fontdimen8\textfont3}}}%
   \mkern-4mu\hbox{\usefont{U}{lasy}{m}{n}\symbol{41}}}}

\begin{document}

\input epsf
\title{Transport parameters from AMS-02 F/Si data\\ and fluorine source abundance}

\author{E. Ferronato Bueno\inst{1}\thanks{\url{e.ferronato.bueno@rug.nl}}
  \and L. Derome\inst{2}
  \and Y. G\'enolini\inst{3}
  \and D. Maurin\inst{2}\thanks{\url{david.maurin@lpsc.in2p3.fr}}
  \and V. Tatischeff\inst{4}
  \and M. Vecchi\inst{1}
}

\authorrunning{E. Ferronato Bueno et al.}

\institute{
Kapteyn Astronomical Institute, University of Groningen, Landleven 12, 9747 AD Groningen, The Netherlands
\and LPSC, Université Grenoble Alpes, CNRS/IN2P3, 53 avenue des Martyrs, 38026 Grenoble, France
\and LAPTh, Universit\'e Savoie Mont Blanc \& CNRS, Chemin de Bellevue, 74941 Annecy Cedex, France
\and Université Paris-Saclay, CNRS/IN2P3, IJCLab, F-91405, Orsay, France
}

\date{Received / Accepted}

\abstract
{The AMS-02 collaboration recently released cosmic-ray F/Si data of unprecedented accuracy. Cosmic-ray fluorine is predominantly produced by fragmentation of heavier progenitors, while silicon is mostly accelerated at source. This ratio is thus maximally sensitive to cosmic-ray propagation.
}
{We study the compatibility of the transport parameters derived from the F/Si ratio with those obtained from the lighter Li/C, Be/C, and B/C ratios. We also inspect the cosmic-ray source abundance of F, one of the few elements with a high first ionisation potential but only moderately volatile, and a potentially key element to study the acceleration mechanism of cosmic rays.}
{We use the 1D diffusion model implemented in the \usinegen{} code and perform $\chi^2$ analyses accounting for several systematic effects (energy correlations in data, nuclear cross sections and solar modulation uncertainties). We also take advantage of the \exfor{} nuclear database to update the F production cross sections for its most important progenitors (identified to be $^{56}$Fe, $^{32}$S, $^{28}$Si, $^{27}$Al, $^{24}$Mg, $^{22}$Ne, and $^{20}$Ne).}
{The transport parameters obtained from AMS-02 F/Si data are compatible with those obtained from AMS-02 (Li,Be,B)/C data. The combined fit of all these ratios leads to a $\chimindof{}\approx 1.1$, with $\lesssim 10\%$ adjustments of the B and F production cross sections (the latter are based on very few nuclear data points, and would strongly benefit from new measurements). The F/Si ratio is compatible with a pure secondary origin of F, with a best-fit relative source abundance ${\rm (^{19}F/^{28}Si)}_{\rm CRS}\sim 10^{-3}$ and an upper limit of $\sim 5\times 10^{-3}$. Unfortunately, this limit is not sufficient to test global acceleration models of cosmic-ray nuclei, for which values at the level of $\sim 10^{-4}$ are required. Such levels could be attained with F/Si data of a few percent accuracy at a few tens of TV, possibly within reach of the next generation of cosmic-ray experiments.
}
{}


\maketitle

\section{Introduction}

The AMS-02 (Alpha Magnetic Spectrometer) experiment, on the International Space Station since 2011, has the capability to identify cosmic-ray (CR) elements up to Ni, and it has already provided a huge body of high-precision CR data \citep{2021PhR...894....1A}. In particular, AMS-02 provides elemental fluxes and ratios at percent level precision from $\sim 1$~GV to a few TV in rigidity ($R=pc/Z$). These data challenge the standard picture \citep[e.g.,][]{2007ARNPS..57..285S,2015ARA&A..53..199G,2019IJMPD..2830022G} of CR transport in the Galaxy.

Broadly speaking, CR nuclei can be separated in two categories, related to the process making up most of their flux: primary species mostly come from material synthesised and later on accelerated in astrophysical sources \citep{1997ApJ...487..182M,1997ApJ...487..197E,2019ApJS..245...30L}, while secondary species originate from nuclear interactions (fragmentation) of the primary species on the interstellar medium (ISM) \citep{1964ocr..book.....G}.
In releasing their data, the AMS-02 collaboration started from the most abundant species, namely the light primary species H \citep{2015PhRvL.114q1103A} and He \citep{2015PhRvL.115u1101A}, then move on to C and O \citep{2017PhRvL.119y1101A}, and more recently, to Ne, Mg, Si \citep{2020PhRvL.124u1102A}, and also Fe \citep{2021PhRvL.126d1104A}. These data are crucial for the modelling of secondary species, because the break-up of these nuclei populate and contribute, to varying fractions, to the flux of all species lighter than Si. Examples of mixed species are N, Na and Al \citep{2021PhRvL.127b1101A}, which receive a roughly similar amount of primary and secondary contributions at a few GV \citep{2020CoPhC.24706942M}.
Secondary-to-primary ratios are of special interest, since they allow to study the transport parameters independently of source ones \citep[e.g.,][]{1964ocr..book.....G,2002A&A...394.1039M,2015A&A...580A...9G}. The B/C ratio was the first secondary-to-primary ratio released by the AMS-02 collaboration \citep{2016PhRvL.117w1102A}, followed by Li/C and Be/C \citep{2018PhRvL.120b1101A} and by the $^3$He/$^4$He isotopic ratio \citep{2019PhRvL.123r1102A}.

We focus in this paper on the recently released AMS-02 F/Si ratio \citep{2021PhRvL.126h1102A}, extending our previous efforts to interpret AMS-02 secondary-to-primary ratios \citep{2017PhRvL.119x1101G,2019PhRvD..99l3028G,2020A&A...639A.131W,2022A&A...668A...7M}. While the F/Si ratio can be considered as a secondary-to-primary ratio (see below), it has not been recognised as such, except in one very recent study \citep{2022ApJ...925..108B}. Indeed, the bulk of the CR modelling literature focused on the B/C ratio \citep{1998ApJ...509..212S,2010A&A...516A..66P,2011ApJ...729..106T,2012A&A...544A..16T,2014PhRvD..89d3013C,2015JCAP...10..034K,2019PhRvD..99l3028G,2020PhRvD.101b3013E}, the $^3$He/$^4$He ratio \citep{1994ApJ...431..705S,1997AdSpR..19..817W,2012A&A...539A..88C,2012Ap&SS.342..131T,2017ICRC...35..210P,2019PhLB..789..292W}, and the sub-Fe/Fe ratio\footnote{In most publications, Sub-Fe=Sc+Ti+V, i.e. $Z=21-23$ elements.} \citep{1987ApJS...64..269G,1992ApJ...390...96W,2001ApJ...547..264J,2001ApJ...555..585M,2008JCAP...10..018E}. 
The reason behind this oversight can hardly be explained by the quality of the data: previous detectors measuring F and Si, e.g. HEAO3 (High Energy Astrophysical Observatory) \citep{1990A&A...233...96E} and ACE (Advanced Composition Explorer) \citep{2009ApJ...698.1666G,2013ApJ...770..117L}, had similar systematics over a wide range of charges, with only $\sim\!\sqrt{2}$ larger statistical uncertainties for F/Si compared to sub-Fe/Fe (Si/Fe~$\sim1.5$ and F/Sub-Fe~$\sim 0.8$, \citealt{1990A&A...233...96E}).
But it is possibly related to nuclear physics: until recent times, the community was lacking reliable nuclear production cross sections for F. As we will see, nuclear data for its production are particularly scarce, with only a few relevant measurements in the 90's and 00's. Moreover, the production of F involves more progenitors compared to other secondary species, which means than the number of cross sections required for CR calculations is larger. As discussed below, the main five progenitors of $^{19}$F\footnote{We use indifferently F or $^{19}$F (only stable isotope for this element).} ($^{20,22}$Ne, $^{24}$Mg, $^{28}$Si, and $^{56}$Fe) only make $\sim 60\%$ of the total production. In comparison \citep{2012A&A...539A..88C,2018PhRvC..98c4611G,2022A&A...668A...7M}, only two progenitors ($^{12}$C and $^{16}$O) make $\sim 60\%$ of all the Li, Be, and B, and even one single progenitor, $^4$He (resp. Fe), makes $\sim 90\%$ of $^3$He (resp. sub-Fe).

A possible issue with using F/Si as a probe of CR transport is that F is expected to have a tiny but non-zero primary abundance. Indeed, whereas the ratio of $^3$He, Li, Be, B, and sub-Fe to their progenitors are in trace amount in the Solar system (SS), F/Si is measured at the percent level \citep{2003ApJ...591.1220L}: this should translate very roughly into a percent level contribution in the F/Si CR flux ratio at GeV/n, unless some specific mechanism were to accelerate F more efficiently than Si. Actually, in the range of elements measured by AMS-02, F might be a key species to study the origin of cosmic rays \citep{2019ApJS..245...30L,2021MNRAS.508.1321T}. As a matter of fact, two competing explanations have been advocated to explain the CR source composition derived from CR data, either from a first ionisation potential (FIP) bias---similar to that found in solar energetic particles \citep{Meyer1979,Meyer1985}---or a volatility bias \citep{1997ApJ...487..197E}. The two properties are correlated for most elements, but F is one of the few that breaks this pattern \citep{1997ApJ...487..182M}. In the light of the new AMS-02 data, this makes the study of the F source abundance worth exploring.

The paper is organised as follows: in Sect.~\ref{sec:setup}, we recall the transport equation and describe the propagation model and configuration used in our analysis.
In Sect.~\ref{sec:standaloneFSi}, we inspect the constraints AMS-02 F/Si data set on the transport parameters (assuming F is a pure secondary species) and check their compatibility with the constraints set by the combined AMS-02 (Li,Be,B)/C data. In Sect.~\ref{sec:B/C+F/Si}, we analyse the behaviour of a combined analysis of the above-two datasets. In Sect.~\ref{sec:Fsource}, we draw constraints on the abundance of F (relative to Si) in CR sources, discussing whether this is consistent with the expectations of current CR source composition modelling. We conclude in Sect.~\ref{sec:conclusions}.
For readability, we postpone to Appendices crucial but more technical details: in App.~\ref{app:methodo}, we describe the F/Si fitting strategy, which involves priors on Solar modulation and nuclear cross-section parameters, and also the covariance matrix of uncertainties for AMS-02 F/Si data; in App.~\ref{app:prod_xs}, we identify the most important progenitors of F and rescale to recent nuclear data their production cross sections; in App.~\ref{app:prim}, we discuss how well our model reproduce primary CR data, and whether the results discussed in the main text are sensitive to the assumption made on the CR source spectral shape.

\section{Model and free parameters}
\label{sec:setup}

We use here the same model and methodology as used in our previous analyses \citep{2019PhRvD..99l3028G,2020PhRvR...2b3022B,2020A&A...639A..74W,2020A&A...639A.131W}, and we refer the reader to these papers for complementary information. We provide below a summary description of the model and of the ingredients relevant for the analysis of F/Si data.

\subsection{Transport equation in 1D model}
\label{sec:model}

Assuming isotropy, the flux of a CR ion is related to the differential density, $dN/dE\equiv N$ (for short), by $\psi=vN/(4\pi)$.
The differential density $N^k$ for a CR species $k$ is calculated from the transport equation \citep{1990acr..book.....B}. For homogeneous and isotropic diffusion (coefficient $K$) and constant convective transport ($V_c$) taken perpendicular to the Galactic plane (symmetric w.r.t. the disc), the steady-state equation in the thin-disc approximation (e.g., \citealt{1992ApJ...390...96W}) becomes \citep{2010A&A...516A..66P}
\begin{equation}
\label{eq:1D}
\begin{split}
  &\left(  -K \frac{\partial^2}{\partial z^2}
 +V_c \frac{\partial }{\partial z} + \frac{1}{\gamma\,\tau^k_{\rm rad}} + 2h\,\delta(z) \sum_{t\,\in\rm ISM} n_t\,v\,\sigma^{k+t}_{\rm inel}\right) N^k \\
 & \hspace{2mm}+\;2h\,\delta(z)\frac{\partial}{\partial E}\left( b^k N^k - c^k \frac{\partial N^k}{\partial E}\right)\\
 & \hspace{2mm}=  2h\,\delta(z) Q^k(E) + \frac{N^r}{\gamma\,\tau^{r\to k}_{\rm rad}} + 2h\,\delta(z) \!\sum_{t\,\in\rm ISM}\!\sum_p\! n_t\,v\,\sigma^{p+t\to k}_{\rm prod}N^p\,,
\end{split}
\end{equation}
with
\begin{eqnarray}
\label{eq:b}
b(E)\!&\!=\!&\! \Big\langle\frac{dE}{dt}\Big\rangle_{\rm ion,\,coul.} 
   - \frac{\vec{\nabla}.\vec{V}}{3} E_k\left(\frac{2m+E_k}{m+E_k}\right)
 + \frac{(1+\beta^2)}{E}\, K_{pp},\\
\label{eq:c}
c(E)\!&\!=\!&\!  \beta^2 \, K_{pp},
\end{eqnarray}
and
\begin{equation}
K_{pp}\times K= \frac{4}{3}\;V_a^2\;\frac{p^2}{\delta\,(4-\delta^2)\,(4-\delta)}\,.
\label{eq:KppK}
\end{equation}

The various quantities appearing in Eq.~(\ref{eq:1D}) are: (i) the half-life $\tau_{\rm rad}$, either leading to a disappearance if CR species $k$ is unstable, or to a radioactive source term (if $r$ decays into $k$); (ii) nuclear reaction rates $\sum_t n_tv\,\sigma$ on the interstellar medium (ISM) targets $t$ of density $n_t$, either associated to a net loss with a destruction cross section $\sigma_{\rm inel}^{k+t}$, or to a secondary source term, summed over $p$ progenitors (heavier than $k$) with a production cross section $\sigma_{\rm prod}^{p+t\to k}$ (straight-ahead approximation is assumed); (iii) continuous energy losses and gains in the disc only($b$ and $c$ terms), accounting for ionisation and Coulomb losses \citep{1994A&A...286..983M,1998ApJ...509..212S}, adiabatic expansion in the Galactic wind, and first and second order contribution from reacceleration; (iv) the so-called primary source term, $Q^k(E)$, from astrophysical sources in the thin disc. Finally, Eq.~(\ref{eq:KppK}) links the diffusion coefficient is momentum $(K_{pp})$ and in space $(K)$ from a minimal reacceleration model \citep{1988SvAL...14..132O,1994ApJ...431..705S}, via the Alfv\'enic speed $V_a$.

\subsection{Propagation code, setup, and ingredients}

The above equation couples about a hundred CR species (for $Z<30$) over a nuclear network of more than a thousand reactions. To solve this triangular matrix of equations, one possibility is to start from the heavier nucleus, which is always assumed to be a primary species, and then to proceed down to the lightest one. 
In the process, isotopic source abundances are fixed to Solar system ones \citep{2003ApJ...591.1220L}, but elemental abundances are normalised to a high energy CR data point. However, accounting for the anomalous isotopic overabundance of $^{22}$Ne/$^{20}$Ne at source \citep[e.g.,][]{2018ARNPS..68..377T} is mandatory not to bias the F calculation (see App.~\ref{app:anomaly}). For this reason, we also rescale the isotopic abundances in order to match a low-energy ACE-CRIS (Cosmic Ray Isotope Spectrometer) $^{22}$Ne/$^{20}$Ne data point \citep{2005ApJ...634..351B}.

All the results derived in this paper are based on the \usine{}\footnote{\url{https://lpsc.in2p3.fr/usine}} public code \citep{2020CoPhC.24706942M}, which provides a full implementation of the 1D model and its solutions.
This model assumes an infinite slab with a thin disc and a thick halo \citep{2001ApJ...547..264J}: the gas and sources (with energy losses and reacceleration) are in the disc, while a spatially-independent diffusion and constant convection are enabled in the diffusion halo; CRs freely escape at the boundary $L$ of the halo. In this model, CR fluxes only depend on the vertical coordinate, and the thin disc approximation allows to solve analytically Eq.~(\ref{eq:1D}) in the halo; the energy part in the disc is solved numerically. Despite its simplicity, this model captures all the essential features of CR transport \citep{2001ApJ...547..264J}, as further illustrated by the occasional use of this model by developers of the numerical code {\tt DRAGON} \citep{2018JCAP...07..006E} in some of their studies \citep{2020PhRvD.101b3013E,2021PhRvD.103l3010S}.

As in our previous publications, we fix the disc half-thickness $h=100$~pc and the gas density $n_{\rm ISM}=1~{\rm cm}^3$ ($90\%$ H and $10\%$ He in number) to recover the local gas surface density \citep{2001RvMP...73.1031F}. The flux of stable species is independent of $K_0/L$ \citep[e.g.,][]{2001ApJ...555..585M}, so that we can fix the halo size without loss of generality. We use here $L=5$~kpc, a value consistent with the latest determination from AMS-02 data \citep{2020A&A...639A..74W,2022A&A...667A..25M}.
For the source term $Q(R)$, we assume a simple power-law in rigidity with a universal slope. This was sufficient to give an excellent match (no fit) to AMS-02 C, N, and O data in our (Li,Be,B)/C analyses \citep{2019PhRvD..99l3028G,2020A&A...639A.131W}. It also gives a fair match to the flux of the main progenitors of F, and similarly to the B/C ratio, this implies that the F/Si ratio becomes independent of its progenitors' source spectrum \citep{2002A&A...394.1039M,2011A&A...526A.101P,2015A&A...580A...9G}. We check in App.~\ref{app:prim} that the use of broken-power laws (instead of a simple power-law), while better fitting the primary flux data, has its own issues. In any case, the results presented below are insensitive to this choice.

\subsection{\BIG{}, \SLIM{}, and \QUAINT{} configurations}
Besides $V_a$ and $V_c$ (for reacceleration and convection), the free parameters of our analysis are related to the diffusion coefficient, parameterised as a phenomenologically motivated power-low with a break at both low-rigidity \citep{2019PhRvD..99l3028G,2019PhRvD.100d3007V,2020A&A...639A.131W} and high-rigidity \citep{2012ApJ...752L..13T,2017PhRvL.119x1101G,2018JCAP...01..055R,2019PhRvD..99l3028G,2020JCAP...01..036N}:
\begin{equation}
  \label{eq:def_K}
  K(R) = {\beta^\eta} K_{0} \;
  {\left\{ 1 \!+ \left( \frac{R_l}{R} \right)^{\frac{\delta-\delta_l}{s_l}} \right\}^{s_l}}
  {\left\{  \frac{R}{R_0\!=\!1\,{\rm GV}} \right\}^\delta}\,
  {\left\{  1 \!+ \left( \frac{R}{R_h} \right)^{\frac{\delta-\delta_h}{s_h}}
    \right\}^{-s_h}}\!\!\!\!\!\!.
\end{equation}
As discussed in \citet{2019PhRvD..99l3028G}, the breaks delineate specific low, intermediate, and high-rigidity ranges that can be related to both features in the data \citep{2013ApJ...770..117L,2016ApJ...831...18C,2017arXiv171202818W,2019SciA....5.3793A,2020PhRvL.125y1102A,2021PhR...894....1A} and peculiar microphysics mechanisms \citep{2004ApJ...614..757Y,2006ApJ...642..902P,2010ApJ...725.2110S,2012PhRvL.109f1101B,2014ApJ...782...36E,2018PhRvL.121b1102E,2021MNRAS.tmp..378F}.

We re-use here the three propagation scenarios proposed in \citet{2019PhRvD..99l3028G}\footnote{We stress that there is a rather long list of additional effects or alternative modelling that may be relevant to interpret CR fluxes, in particular regarding features at high energy \citep{2012ApJ...752...68V}: energy-dependence of the nuclear cross sections \citep{2015ApJ...811...11K}, source spectral breaks or diversity of CR sources \citep{2013ApJ...763...47P,2015MNRAS.447.2224B,2016PhRvD..93h3001O,2018MNRAS.474L..42R}, stochasticity of the sources \citep{2012A&A...544A..92B,2012MNRAS.421.1209T,2013MNRAS.435.2532T,2015RAA....15...15L,2017A&A...600A..68G,2017PhRvD..96b3006L}, reacceleration \citep{1987ApJ...316..676W,2014A&A...567A..33T}, or secondary production at source \citep{2003A&A...410..189B,2009PhRvL.103h1103B,2012A&A...544A..16T,2014PhRvD..89d3013C,2014PhRvD..90f1301M,2021PhRvD.104j3029M}, single-source hypothesis \citep{2015PhRvL.115r1103K,2018PhRvD..97f3011K}, etc. These effects might be relevant to interpret the DAMPE data \citep{2019SciA....5.3793A} at a few tens of TeV \citep[e.g.,][]{2020ApJ...903...69F,2021ApJ...911..151M}. However, none of these affect are considered here, as we wish to test whether a more standard model can already explain the very precise AMS-02 data.}, namely \BIG{}, \SLIM{}, and \QUAINT{}: (i) \BIG{} enables the `full' transport complexity, i.e. diffusion ($K_0,\,\delta,\,R_l,\,\delta_l,\,s_l,\,R_h,\,\delta_h,\,s_h$, but $\eta\,$=1), convection ($V_c$), and reacceleration ($V_a$); (ii) \SLIM{} is a minimal diffusion scenario with $\eta\,$=1 and $V_c\,$=$\,V_a\,$=\,0; (iii) \QUAINT{} is a more standard diffusion/convection/reacceleration scenario, i.e. no low-energy break ($R_l\,$=0) but possible sub-relativistic upturn of the diffusion coefficient via $\eta$. The latter two scenarios can be viewed as two different limiting regimes of \BIG{}. All these scenarios lead to similar predictions in the high-energy regime ($\gtrsim10$~GV), and they all give a very good fit to light secondary-to-primary data \citep{2020A&A...639A.131W} and also antiprotons \citep{2020PhRvR...2b3022B}.
To limit the number of free parameters in our analysis, we assume a fast transition $s_l$ = 0.04 for the low-rigidity break and also fix the three high-rigidity break parameters ($R_h,\delta_h,s_h$) to the values in \citet{2019PhRvD..99l3028G}; as showed by these authors, this choice only marginally affects the remaining parameters. 

\subsection{Outline of the F/Si fit procedure}

The need for an improved methodology to interpret high-precision AMS-02 data was discussed at length in \citet{2019A&A...627A.158D}, and  we follow it here. We perform a $\chi^2$ analysis (App.~\ref{app:chi2_def}), accounting for Solar modulation and nuclear cross-sections uncertainties via nuisance parameters (Apps~\ref{app:nui_phiff} and \ref{app:nui_xs}), in order not to bias the determination of the transport and source parameters; relevant cross sections for the production of F are updated based on recent nuclear data (App.~\ref{app:rescaling}). We also account for the covariance matrix of data systematic uncertainties (App.~\ref{app:cov_mat}).

\begin{table}[t]
\caption{List of ingredient (used for the analysis) and their description.}
\label{tab:definitions}
\centering
{
\footnotesize
\begin{tabular}{ll}
\hline\hline
Name                 & Description                                                                  \\
\hline
\multicolumn{2}{l}{\em Free transport parameters (in particular Eq.~\ref{eq:def_K})}                \\
$K_0$                & $K(R)$ normalisation\tablefootmark{a} in [kpc$^2$~Myr$^{-1}$]                \\
$\delta$             & Diffusion slope (at intermediate energies)                                   \\
$\delta_l$, $R_l$    & Diffusion slope break (at low rigidity $R_l$)                                \\
$\eta$               & Low-energy modification of $K(R)$, via $\beta^\eta$                          \\
$V_c$                & Convection in [m/s] (perpendicular to the disc)                              \\
$V_a$                & Reacceleration in [m/s] (in the thin disc)                                   \\[4mm]
\multicolumn{2}{l}{\em Propagation model configuration (from \citealt{2019PhRvD..99l3028G})}        \\
\SLIM{}              & Free params $= K_0$, $\delta$, $R_l$, $\delta_l$ (main analysis)             \\
\QUAINT{}            & Free params $= K_0$, $\delta$, $\eta$, $V_c$, $V_a$                          \\
\BIG{}               & Free params $= K_0$, $\delta$, $R_l$, $\delta_l$, $V_c$, $V_a$               \\[4mm]
\multicolumn{2}{l}{\em Free source parameters (used in Sect.~\ref{sec:Fsource})}                    \\
$q_{^{19}\rm F}/q_{^{28}\rm Si}$ & Abundance of $^{19}$F relative to $^{28}$Si                      \\[4mm]
\multicolumn{2}{l}{\em Ingredients of $\chi^2$ analysis (see App.~\ref{app:methodo})}               \\[0mm]
                     & {\em [Nuisance parameters (mean and dispersion)]}                            \\
$\mu_\phi$, $\sigma_\phi$ & For Solar modulation level: App.~\ref{app:nui_phiff} and Eq.~(\ref{eq:phi_ff}) \\
$\mu_{\rm XS}$, $\sigma_{\rm XS}$ & For cross-section proxies: App.~\ref{app:nui_xs} and Eqs~(\ref{eq:NSS_Norm}-\ref{eq:NSS_Slope})\!\!\!\\[1mm]
                     & {\em [Covariance matrix]}                                                    \\
$(C)_{ij}$           & For data uncertainties: App.~\ref{app:cov_mat} and Eq.~(\ref{eq:cov})        \\
$\ell_\alpha$        & Correlation length for uncertainty type $\alpha$: Eq.~(\ref{eq:cov})         \\[1mm]
                     & {\em [$\chi^2$ interpretation (best fit)]}                                   \\
\chimindof{}         & $\chi^2$ per degree of freedom ($\lesssim 1$ for good fit): Eq.~(\ref{eq:chi2}) \\
\chipernui{}         & Contrib. of nuisance dof in $\chi^2$ ($0$ if unused): Eq.~(\ref{eq:chi2nuis})\!\!\!\\[4mm]
\multicolumn{2}{l}{\em Cross-section datasets (updated on nuclear data, App.~\ref{app:rescaling})}  \\
\optxii{}            & Updated from the eponymous \galprop{} dataset                                \\
\optxxii{}           & Updated from the eponymous \galprop{} dataset                                \\
\optxiiupxxii{}\!\!\!\!\!\!& Mixture (believed to provide the best description)                     \\
\hline
\end{tabular}
}
\tablefoot{
\tablefoottext{a}{We recall that $1$~kpc$^2$~Myr$^{-1}\simeq 3\cdot 10^{29}$cm$^2$~s$^{-1}$.}
}
\end{table}

To ease the reading of the paper and the navigation between the different names and configurations explored in our analyses, they are gathered and described in Table~\ref{tab:definitions}. The core of our analyses and results presented below relies on the \SLIM{} propagation configuration (pure diffusion configuration). The configurations \QUAINT{} and \BIG{} (with convection and reacceleration) are used to check that our conclusions are insensitive to this choice.
In the following, we present our results based on the updated (for F) production cross-section sets \optxii{}, \optxxii{}, and \optxiiupxxii{} (see App.~\ref{app:rescaling}). The latter set is considered to be the most relevant to better describe the nuclear and CR data (see \citealt{2022A&A...668A...7M} for details).

\section{Analysis of F/Si: consistency with Li,Be,B/C?}
\label{sec:standaloneFSi}

In this section, we show our results for the propagation configuration \SLIM{}, whose parameters are listed in Sect.~\ref{sec:model}. Similar conclusions are obtained for the \BIG{} and \QUAINT{} configurations.

\begin{table*}[t]
\caption{Comparisons of best-fit transport parameters (in \SLIM{}) from various combinations of AMS-02 data.}
\label{tab:best_fit}
\centering
{
\footnotesize
\begin{tabular}{llcccccccc}
\hline\hline
Config.          &   Ref.    & $\delta$ & $\displaystyle\log_{10}\left(\frac{K_0}{\mathrm{1\,kpc^2\,Myr^{-1}}}\right)$ & $R_l$ [GV]  &$\delta_l$ &  $\displaystyle\log_{10}\left(\frac{q_{^{19}\rm F}}{q_{^{28}\rm Si}}\right)$ & \chimindof{}&\chipernui{}\\
\hline
F/Si~~($q_{^{19}\rm F}=0$)   &this paper & $0.449\pm0.020$ & $-1.156\pm0.054$ & $4.21\pm0.33$ & $-0.45\pm0.33$ & -     & 0.68 & 0.04 \\
B/C              & [Gé19]    & $0.509\pm0.017$ & $-1.361\pm0.047$ & $4.38\pm0.22$ & $-0.87\pm0.32$ & -     & 0.98 & 0.04 \\
(Li,Be,B)/C      & [Ma22]    & $0.500\pm0.010$ & $-1.330\pm0.025$ & $4.67\pm0.40$ & $-0.36\pm0.23$ & -     & 1.21 & 0.50 \\
F/Si+(Li,Be,B)/C &this paper & $0.492\pm0.008$ & $-1.293\pm0.021$ & $4.55\pm0.24$ & $-0.39\pm0.13$ & $-3.0^{+0.6}_{-\infty}$ & 1.12 & 0.4 \\
\hline
\end{tabular}
}
\tablefoot{The data combinations (line entries) analysed are : F/Si alone with \optxiiupxxii{} cross-section set (this analysis), B/C alone \citep{2019PhRvD..99l3028G}, combined (Li,Be,B)/C with \optxiiupxxii{} cross-section set \citep{2022A&A...668A...7M}, and combined (Li,Be,B)/C and F/Si data with updated production cross sections (App.~\ref{app:prod_xs}). See text for discussion.}
\end{table*}

\subsection{F/Si vs B/C}
\begin{figure}[t!]
   \centering
   \includegraphics[width=\columnwidth]{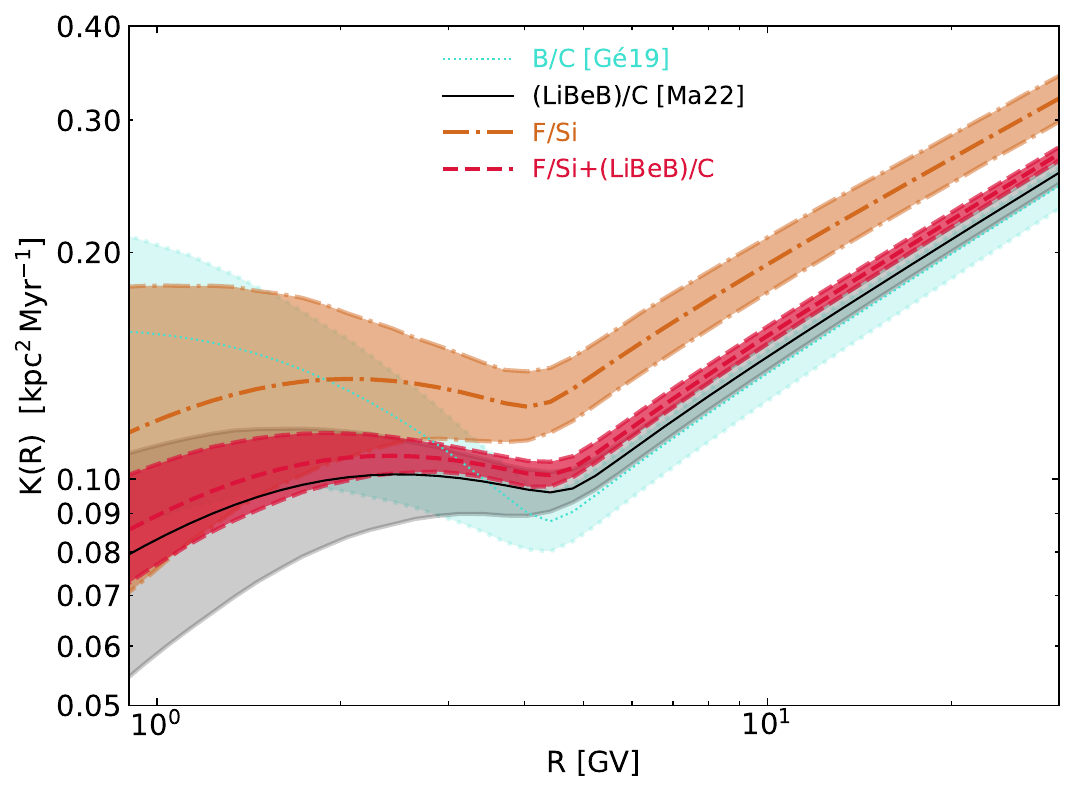}
   \caption{Best-fit and 1$\sigma$ envelopes for the diffusion coefficient, Eq.~(\ref{eq:def_K}), for different secondary-to-primary ratio data combinations. The corresponding parameters are gathered in Table~\ref{tab:best_fit}.
   \label{fig:KR}}
\end{figure}

Table~\ref{tab:best_fit} shows the best-fit parameters obtained from various combinations of secondary-to-primary data, as obtained from this analysis or from some of our previous publications (all relying on the methodology recalled above and in App.~\ref{app:methodo}). The first two lines show the results obtained from the analysis of a single secondary-to-primary ratio, namely F/Si (this paper) and the widely used B/C \citep{2019PhRvD..99l3028G}. The last two columns show that both the B/C and F/Si fits are excellent ($\chimindof{}\lesssim1$), without using the production cross-section degrees of freedom ($\chipernui{}\approx 0$). Both ratios lead to consistent values for $R_l$ and $\delta_l$ (low-rigidity break) and give $\delta\approx 0.5$ consistently (at $2\sigma$ level). This means that considering F as a pure secondary species is a fair approximation, and that F/Si confirms the trend, in the diffusion coefficient Eq.~(\ref{eq:def_K}), for a low-rigidity break \citep{2019PhRvD..99l3028G,2019PhRvD.100d3007V,2020A&A...639A.131W} and Kraichnan-like slope $\delta$ \citep{2019PhRvD..99l3028G,2020A&A...639A.131W,2021PhRvD.103j3016K,2022PhRvD.105j3033K}.
The uncertainties on the parameters are slightly larger for the F/Si analysis than for the B/C analysis,
translating into larger $1\sigma$ contours of the associated reconstructed $K(R)$: compare the dotted magenta and dash-dotted orange contours for the B/C and F/Si fits respectively in Fig.~\ref{fig:KR}.
This difference is related to the less abundant F and Si fluxes compared to the B and C ones, with larger statistical uncertainties for the AMS-02 F/Si data compared to B/C ones (especially at high rigidity).
We note, though, that the best-fit value of the diffusion coefficient normalisation, $K_0$, differs significantly between the F/Si and B/C analyses. However, as highlighted in \citet{2020A&A...639A.131W}, considering a single ratio could lead to biased results on $K_0$, because the latter is mostly degenerate with the production cross-section nuisance parameters. We come back to this difference in Sect.~\ref{sec:B/C+F/Si}.

\subsection{F production uncertainties and F/Si data systematics}
\begin{figure}[t]
\includegraphics[width=\columnwidth]{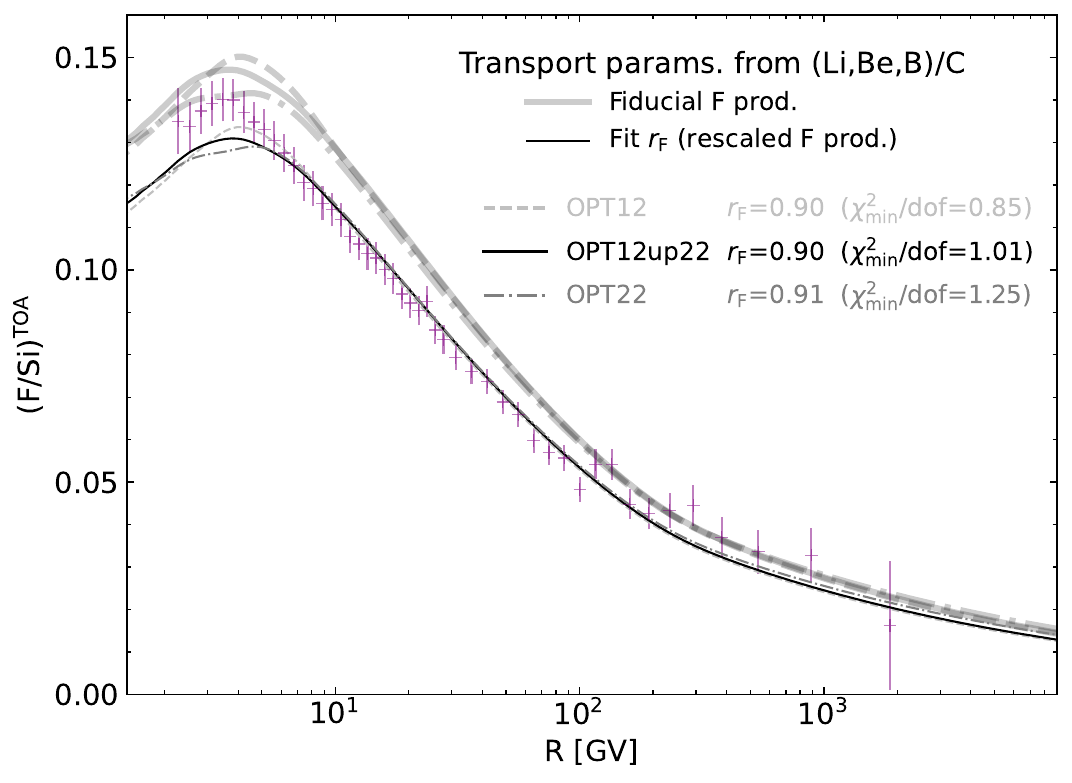}
  \caption{Comparison of F/Si data \citep{2021PhRvL.126h1102A} with calculations (\SLIM{} propagation configuration) calibrated on the (Li,Be,B)/C transport parameters (see Table~\ref{tab:best_fit}). The thick grey lines show the direct calculation based on our three production cross-section sets (dashed line for \optxii{}, solid line for \optxiiupxxii{}, and dotted line for \optxxii{}). The thin lines result from the additional fit of a global factor $r_{\rm F}$ (rescaling the overall production of F) to better match the data. A pure secondary origin of fluorine is assumed in both calculations (see text for discussion).
  \label{fig:FSi_rescaledXS}}
\end{figure}
As an alternative view, we inspect here whether the F/Si ratio predicted from the (Li,Be,B)/C-derived transport parameters \citep{2022A&A...668A...7M} are consistent with the data. This is illustrated in Fig.~\ref{fig:FSi_rescaledXS}, where the thick grey lines show the predicted F/Si ratio (pure secondary hypothesis, not a fit); the purple symbols show the AMS-02 data. The model calculations clearly overshoot the data. By allowing an overall rescaling $r_{\rm F}$ of the F production cross sections, the thin lines shows that the production cross sections---for any of our production sets---must be rescaled by a factor $\approx 0.9$ to match the F/Si data, i.e. consistent with the typical $\sim 10\%$ errors in the nuclear data. This factor is comparable to the scaling applied by \citet{2022ApJ...925..108B}, with the \galprop{} code, to match the F/Si data.

Also highlighted in the legend of Fig.~\ref{fig:FSi_rescaledXS} are the \chimindof{} values obtained after the cross-section rescaling. The \optxiiupxxii{} set is favoured with $\chimindof{}=1.01$, although the model seems significantly offset from  data below 10 GV. We recall that in this range, the `Acc.~norm.' data systematics dominates (see top panel of Fig.~\ref{fig:sub_errors}), with a correlation length of $\ell_{\rm Acc.~norm.}=1$ decade. Such a correlation means that if one data point moves, all the others (over the correlation length) follows. In practice, the model is thus only $\gtrsim 1\sigma$ away from the $n$ low-rigidity points (global normalisation) instead of $n\times 1\sigma$ away ($n$ independent points at $1\sigma$ each). This explains why the fit remains excellent, and moreover, why it is crucial to account for the covariance matrix of the data uncertainties in the analyses. It would actually be extremely useful if the AMS-02 collaboration could directly provide this matrix, as its role is crucial for the interpretation of their data \citep{2019A&A...627A.158D,2020PhRvR...2b3022B,2020PhRvR...2d3017H}.

\section{Combined B/C and F/Si analysis}
\label{sec:B/C+F/Si}

The next step of the analysis is to perform a combined analysis of (Li,Be,B)/C and F/Si data. This allows to also fit the source abundance of $^{19}$F relative to $^{28}$Si, namely $\log_{10}\left(q_{^{19}\rm F}/q_{^{28}\rm Si}\right)$, whose discussion is left to Sect.~\ref{sec:Fsource}.

\subsection{Transport parameters}
The best-fit values of the fit parameters for the \SLIM{} configuration, \optxiiupxxii{} production cross-section, and for the combined analysis of F/Si and (Li,Be,B)/C are reported in Table~\ref{tab:best_fit} (fourth line). For comparison purpose, we also report the results from the (Li,Be,B)/C analysis \citep{2022A&A...668A...7M}. The associated $K(R)$ values and contours are shown in Fig.~\ref{fig:KR}, in red and black respectively. 
From the Table and the figure, we see that the transport parameters are consistent at better than $1\sigma$, and that the combined analysis of the four ratios slightly decreases the uncertainties (compared to the (Li,Be,B)/C analysis). The fit values are actually driven by the subset of (Li,Be,B)/C data, because they combine three times the number of F/Si data, all having slightly smaller statistical uncertainties. Nevertheless, both the \chimindof{} and \chipernui{} values (last two columns in Table~\ref{tab:best_fit}) show an improvement with the fully combined analysis. The results are consistent with no primary source of F (see next section), indicating that $Z=2-5$ and $Z=9$ secondary species can be explained with the same propagation history; the consistency of $^3$He with Li, Be, B data was discussed in \citet{2020A&A...639A.131W}.

\begin{figure}[t]
\includegraphics[width=\columnwidth]{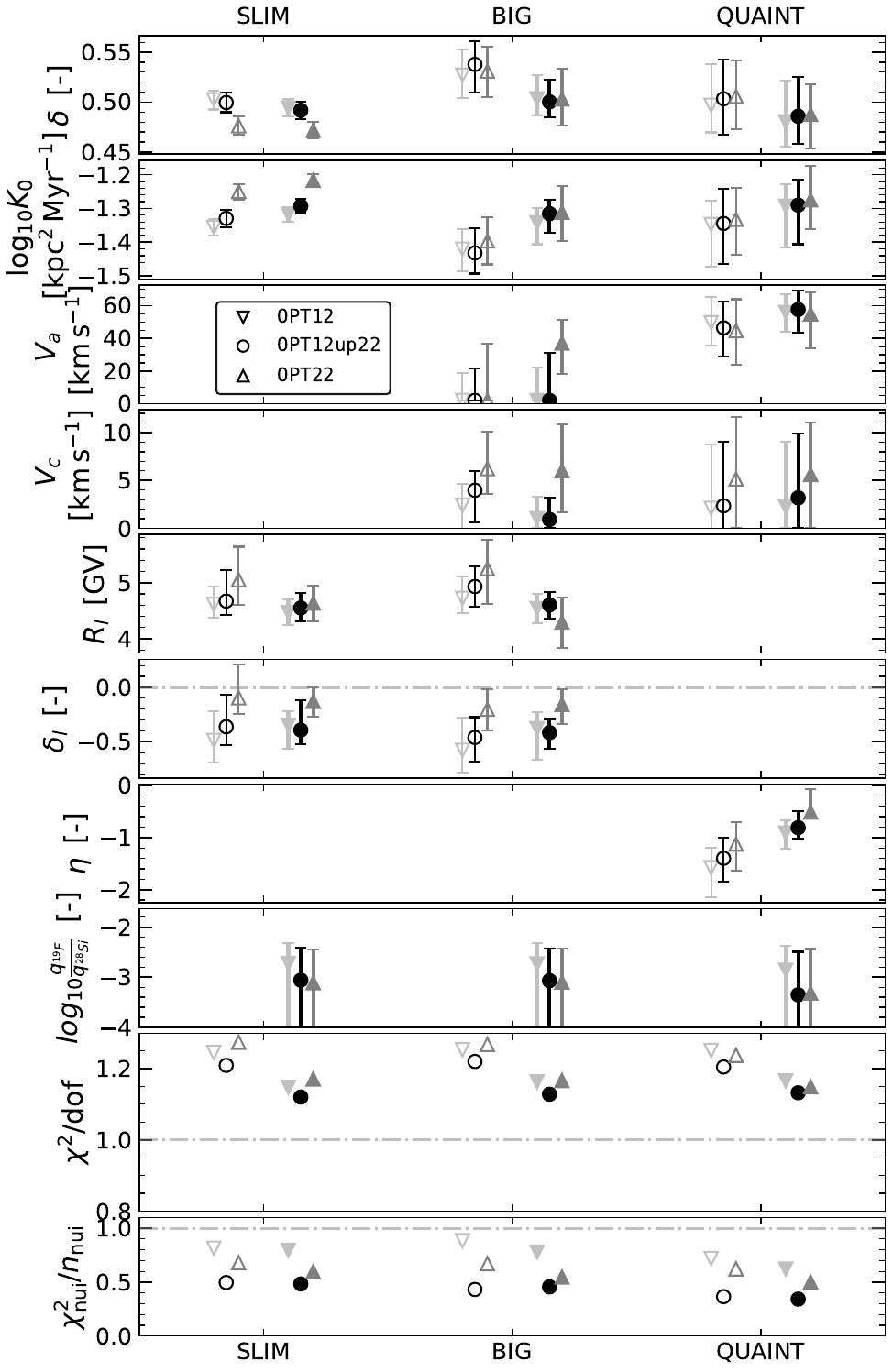}
  \caption{Best-fit parameters and uncertainties from fit of the \SLIM{} (left), \BIG{} (middle) and \QUAINT{} (right) models to AMS-02 data using the cross-section sets \optxii{} (downward silver triangles), \optxiiupxxii{} (black circles), and \optxxii{} (upward grey triangles). The empty symbols with thin error bars correspond to the combined fit to (Li,Be,B)/C data (taken from \citealt{2022A&A...668A...7M}), while the filled symbols with thick error bars correspond to the combined fit to (Li,Be,B)/C and F/Si data (this analysis). See text for discussion.
  \label{fig:model_params}}
\end{figure}
These conclusions are generalised in Fig.~\ref{fig:model_params} for the different transport configurations \BIG{} and \QUAINT{} and different cross-section sets (shown as different symbols). We report the results for the (Li,Be,B)/C \citep{2022A&A...668A...7M} and F/Si+(Li,Be,B)/C analysis (this paper) as empty and filled symbols respectively. The main differences are related to the choice of the production cross-section set, as discussed at length in \citet{2022A&A...668A...7M}.
In particular, for both combination of data, the best \chidof{} is obtained for \optxiiupxxii{}\footnote{We recall that for the (Li,Be,B)/C subset, the significantly larger-than-one \chidof{} was caused by the two low-rigidity Be/B points (upturn) that could not be well fitted by the model \citep{2020A&A...639A.131W}.}.

\subsection{Posteriors for Li, Be, B, and F production cross sections}
\begin{figure}[t]
\includegraphics[width=\columnwidth]{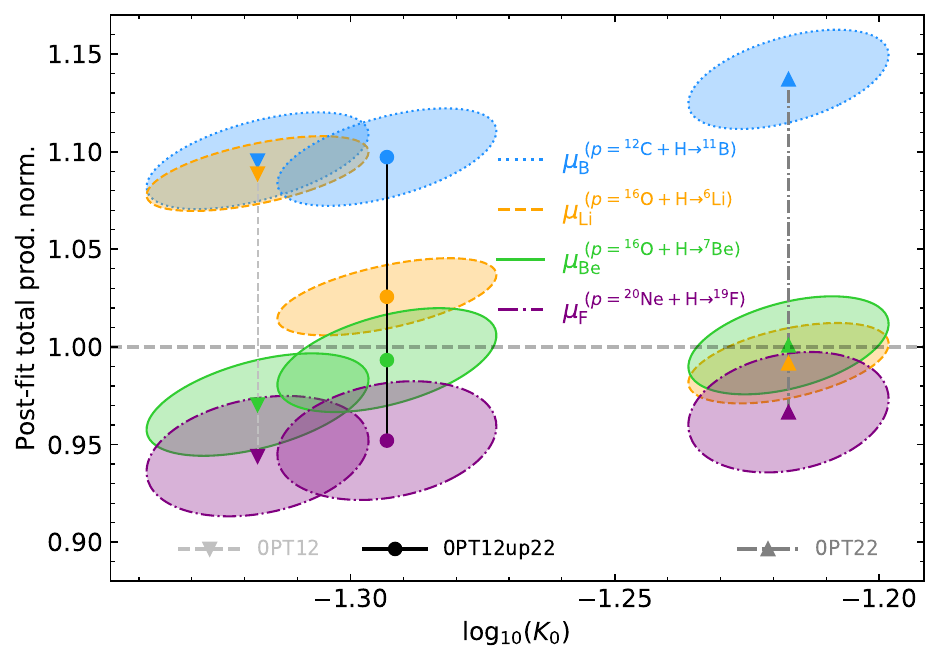}
   \caption{Correlation between log$_{10}(K_{0})$ and the normalisation factor $\mu_Z^{(p)}$, which corresponds to the correction factor applied on the total production cross section (of element $Z$) in order to best-fit AMS-02 F/Si+(Li,Be,B)/C data. The four elements considered are colour-coded: $Z=3$ (Li) in orange, $Z=4$ (Be) in green, $Z=5$ (B) in blue, and $Z=9$ (F) in purple. The $1\sigma$ correlation ellipses are shown for analyses with different cross section sets (in model \SLIM{}): from left to right \optxii{} (filled downward triangles), \optxiiupxxii{} (filled circles), and \optxxii{} (filled upward triangles).
   The horizontal grey dashed line highlights $\mu_Z=1$ (i.e. no modification needed for the production of an element).
\label{fig:ellipses_SLIM}}
\end{figure}

In the F/Si-alone analysis, a lower value of $K_0$ was favoured, compared to that obtained in the  combined analysis. As detailed in \citet{2020A&A...639A.131W} and \citet{2022A&A...668A...7M}, only the combined analysis of secondary-to-primary ratios allows to break the degeneracy between $K_0$ and the production cross sections, by enforcing a unique $K_0$ values for all the ratios. With our methodology (App.~\ref{app:methodo}), in addition to the best-fit parameters, we also get posterior values for the nuisance parameters, i.e. we can inspect the values preferred by the fit for the production cross sections. However, this is not direct as we only use a single proxy reaction per secondary element.
Nevertheless, we can calculate the global normalisation factor $\mu_Z^{(p)}$ required for each secondary element $Z$ to best-fit the data\footnote{The proxy $p$ for the production reaction of an element $Z$ (see Table~\ref{tab:xs_nuis}) only contributes to a fraction $f_p$ of the total. The global bias $\mu_Z^{(p)}$ for the production of this element is calculated from $(\mu_Z^{(p)}-1) = (\mu_p-1)\times f_p$, where $\mu_p$ is the posterior of the normalisation parameter Eq.~(\ref{eq:NSS_Norm}). The relevant $f_p$ are taken from \citet{2018PhRvC..98c4611G}, \citet{2022A&A...668A...7M}, and Fig.~\ref{fig:xs_prodchannels}, with $f_{\mathrm{^{16}O+H\to^{6}Li}}\approx15\%$, $f_{\mathrm{^{16}O+H\to^{7}Be}}\approx19\%$, $f_{\mathrm{^{12}C+H\to^{11}B}}\approx33\%$, and $f_{\mathrm{^{22}Ne+H\to^{19}F}}\approx30\%$.}. We show them in Fig.~\ref{fig:ellipses_SLIM} for our three production cross-section sets.
For Li, Be, and B production, the results are mostly similar to those discussed in Fig.~12 of \citet{2022A&A...668A...7M}, i.e. Li (orange dashed ellipses) is very sensitive to the selected production set, which is not the case for Be and B. Also, Be production is consistent with $\mu_Z\approx1$ (no modification of the production cross sections), but there is a trend for the need to increase the production of B (i.e. $\mu_{\rm B}>1$) and to decrease that of F (i.e. $\mu_{\rm F}<1$). However, the required numbers are of the order of $\sim 10\%$ at most, which is within the range of nuclear uncertainties. New nuclear data are desired in order to confirm this possibility, or to prove that a tension exists between these different flux ratios.

\subsection{Best-matching data}
\begin{figure}[t]
\includegraphics[width=\columnwidth]{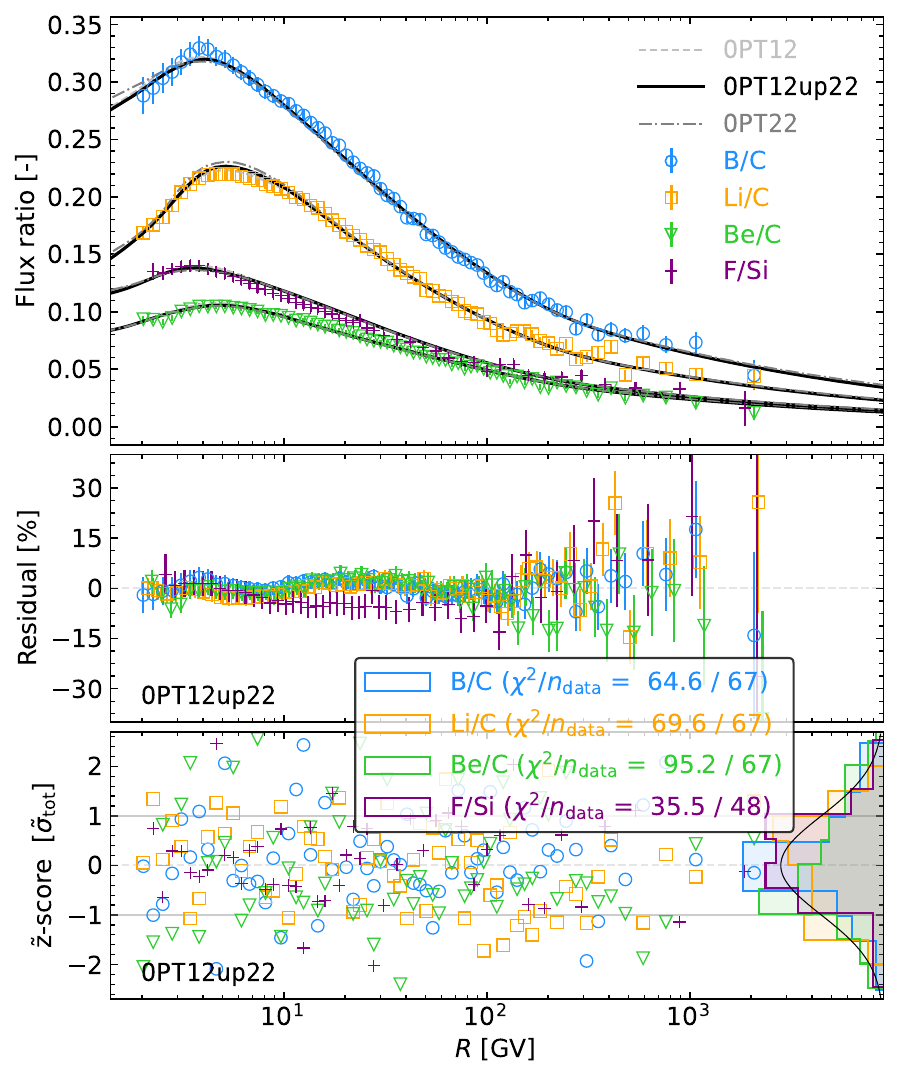}
  \caption{Flux ratios (top), residuals (centre) and $\tilde{z}$-scores (bottom) for B/C (blue circles), Be/C (green downward triangles), Li/C (orange squares), and fluorine (purple upward triangles). The models (top panel) are calculated for the updated \optxii{} (dashed grey line), \optxiiupxxii{} (solid black line), and \optxxii{} (dashed-dotted grey line), from the best-fit transport parameters of the combined analysis of all three species. In the middle and bottom panels, the residuals and $\tilde{z}$-score are shown for the \optxiiupxxii{} configuration only. The distributions in the right-hand side of the bottom panel are histograms of the $\tilde{z}$-score values (projected on the $y$-axis), compared to a $1\sigma$ Gaussian distribution (solid black line). See text for discussion.
   \label{fig:FSiLiBeB_model_vs_data}}
\end{figure}
In this global fit, it is also interesting to look at the detailed match of the model to each ratio, as shown in the top panel of Fig.~\ref{fig:FSiLiBeB_model_vs_data}. 
The middle panel shows the residuals and the bottom panel the $\tilde{z}$-score. The latter is a generalisation of the $z$-score in presence of correlations in the data. We recall that the $z$-score, $z=({\rm model}-{\rm data})/\sigma$, is similar to the residuals but expressed in unit of the total data uncertainties $\sigma_{\rm tot}$. As stressed in \citet{2020PhRvR...2b3022B}, these quantities ignore the correlations in the data, leading to a biased view of the goodness-of-fit between the model and the data. To cure this, \citet{2020PhRvR...2b3022B} proposed to use the $\tilde{z}$-score, which corresponds to the residuals of the eigen vectors (data-model) of the total covariance matrix of data uncertainties (i.e. it accounts for the correlations). By construction, $\chi^2=\sum_i \tilde{z}_i^2$, and the representation of the $\tilde{z}$-score gives an unbiased view of the distance between the model and the data.

In the bottom panel, we also plot on the right-hand side the projected distributions of $\tilde{z}$ ratio by ratio. These distributions are expected to follow a Gaussian distribution of width one (shown as a solid black line) for a good match to the data. More directly, the numbers in the legend give the separate contribution (to the total $\chi^2_{\rm min}$) of these various ratios. We see that they are all in excellent agreement with the data, except for Be/C; the origin of this discrepancy is understood and related to the lowest two data points of this ratio \citep{2020A&A...639A.131W}.

\section{Fluorine source abundance}
\label{sec:Fsource}

The combined F/Si+(Li,Be,B)/C analysis also provides constraints on the relative CR source abundance $(^{19}\rm F/{^{28}\rm Si})_{\rm CRS}$, which are shown in the 8-th panel of Fig.~\ref{fig:model_params} and also reported in Table~\ref{tab:q19F}. These constraints only depend mildly on the production cross-section sets and the transport configuration selected.

From Table~\ref{tab:q19F}, we report a best-fit value of $\sim10^{-3}$. This value is similar to the Solar system ratio ${\rm (^{19}F/^{28}Si)}_{\rm SS}=8.7 \times 10^{-4}$ \citep{Lodders2009}, but should not be compared directly to it (see below). Our fit is also compatible with a null value (i.e. no primary contribution necessary to match the data), and we can draw a 1$\sigma$ upper limit $(^{19}\rm F/{^{28}\rm Si})_{\rm CRS}\sim 5 \times 10^{-3}$.

\begin{table}[t]
\caption{Best-fit values and $+1\sigma$ upper limits (in parenthesis) on the relative source abundance of F.}
\label{tab:q19F}
\centering
{
\footnotesize
\begin{tabular}{lccc}
\hline\hline
                &   \SLIM{}    &    \BIG{}    &  \QUAINT{}  \\
\hline
\optxii{}       & -2.7 (-2.3)  & -2.7 (-2.3)  & -2.8 (-2.4) \\
\optxiiupxxii{} & -3.1 (-2.4)  & -3.1 (-2.4)  & -3.3 (-2.5) \\
\optxxii{}      & -3.1 (-2.4)  & -3.1 (-2.4)  & -3.3 (-2.4) \\
\hline
\end{tabular}
}
\tablefoot{The values are for $\log_{10}\left(q_{^{19}\rm F}/q_{^{28}\rm Si}\right)$ from the analysis of AMS-02 F/Si+(Li,Be,B)/C data. Three propagation configurations (columns) and production data sets (rows) are considered: in all cases, the lower limit for the abundance of $q_{^{19}\rm F}$ is compatible with zero.
}
\end{table}

\subsection{Interpretation}

Fluorine is potentially a key element to distinguish whether the CR source composition is controlled by atomic effects related to the FIP of the elements or by their relative concentration into dust in the interstellar medium \citep{1997ApJ...487..182M}. It is indeed one of the few elements that has a high FIP, $17.4$~eV, close to that of noble gases, but which is only moderately volatile. The F fraction in interstellar dust is not measured, but it is found in non-negligible proportion in some meteorites such as the CI-chondrite Orgueil \citep{Lodders2009}. The equilibrium condensation temperature of F, $T_{\rm cond}=734$~K, is close to that of S, $T_{\rm cond}=664$~K \citep{Lodders2009}, so both elements could be incorporated in about the same proportions in interstellar dust, up to about $20\%$ of their total interstellar abundances \citep{2021MNRAS.508.1321T}. In the model of a preferential acceleration of elements contained in dust \citep{1997ApJ...487..182M,1997ApJ...487..197E}, we can expect F and S to have similar CR source abundances relative to the solar system composition, that is ${\rm [F/Si]}_{\rm CRS} \sim {\rm [S/Si]}_{\rm CRS} = - 0.59$~dex \citep{2021MNRAS.508.1321T}, where ${\rm [X/Y]} = \log_{10}{\rm (X/Y)} - \log_{10}{\rm (X/Y)}_\odot$. Thus, in this model we expect that ${\rm (^{19}F/^{28}Si)}_{\rm CRS} \sim 2.2 \times 10^{-4}$.

In the model where the CR source composition is controlled by a FIP bias similar to that found in solar energetic particles \citep{Meyer1979,Meyer1985}, the source abundance of $^{19}$F can instead be estimated from that of $^{20}$Ne, since the two elements have close masses and FIPs (21.6~eV for Ne). Here, we assume that the $^{19}$F and $^{20}$Ne nuclei in the CR source composition come mainly from the average interstellar medium, and not from the source enriched in Wolf-Rayet wind material at the origin of the $^{22}$Ne excess \citep{2005ApJ...634..351B}. In the FIP model, we thus expect that ${\rm [^{19}F/^{28}Si]}_{\rm CRS} \sim {\rm [^{20}Ne/^{28}Si]}_{\rm CRS} = - 0.76$~dex \citep{2021MNRAS.508.1321T}, and the predicted $^{19}$F abundance relative to $^{28}$Si is ${\rm (^{19}F/^{28}Si)}_{\rm CRS} \sim 1.5 \times 10^{-4}$. Unfortunately, the upper limit on this ratio derived from AMS-02 data is significantly higher than the predicted values in both models and does not allow us to distinguish them.

In their analysis of AMS-02 data, \citet{2022ApJ...925..108B} found an excess below 10~GV in the F spectrum (after reducing their calculated spectrum by about $10\%$) and suggest that it could be explained by a primary F component. The integration of the injection spectra of primary F and $^{28}$Si \citep{2020ApJS..250...27B} above 1~GV, where the excess is reported, gives ${\rm (^{19}F/^{28}Si)}_{\rm CRS} = 2.7 \times 10^{-3}$. If integrated above 0.1~GV (as discussed in \citealp{2022ApJ...925..108B}), the ratio becomes ${\rm (^{19}F/^{28}Si)}_{\rm CRS} = 1.7 \times 10^{-3}$. Both values are consistent with the upper limit obtained in the present work. But they are about an order of magnitude higher than the values predicted from the FIP and condensation temperature of F. Such an overabundance would make F very special compared to all primary and mostly primary CRs from H to Zr, whose abundances are well explained in the model studied by \citet{2021MNRAS.508.1321T}. It seems more likely to us that the excess found in the analysis of \citet{2022ApJ...925..108B} is related to the non-negligible uncertainties in the $^{19}$F production cross sections (see Section~\ref{app:nui_xs}). Note that, yet another explanation of this discrepancy is proposed in \citet{2023PhRvD.107f3020Z}, where the authors use spatially-dependent diffusion.

\subsection{How could we get tighter constraints?}
To go further in the interpretation, we need to improve the present constraints by a factor $\sim 50$.
\begin{figure}[t]
\includegraphics[width=\columnwidth]{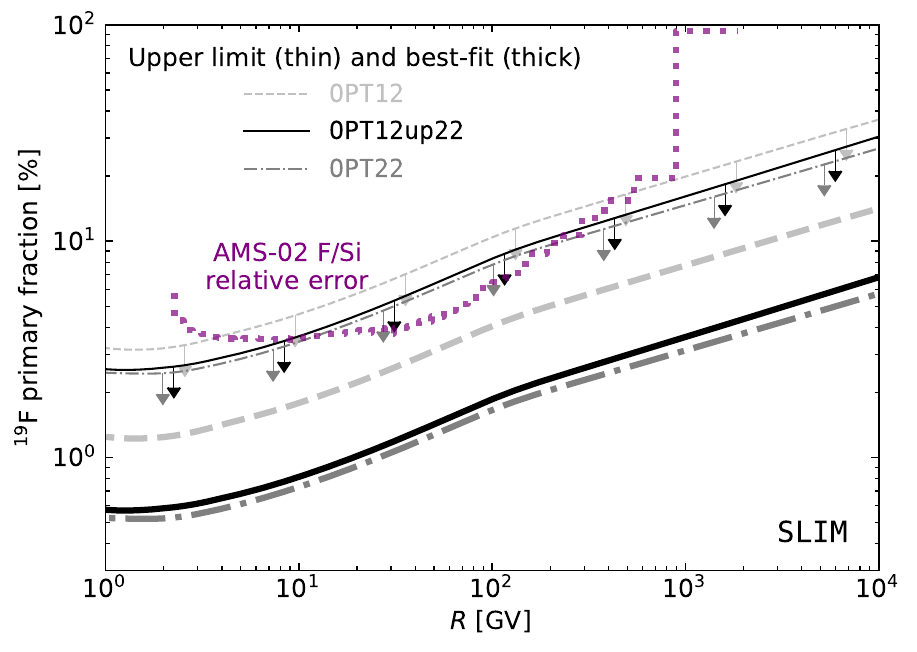}
  \caption{Primary fraction of $^{19}$F as a function of rigidity for the propagation configuration \SLIM{} and the three production cross-section sets \optxii{} (dashed light-grey lines), \optxiiupxxii{} (solid black lines), and \optxxii{} (dash-dotted grey lines). The thick lines are calculated from the best-fit source parameter $q_{^{19}\rm F}$, from the combined analysis of F/Si and (Li,Be,B)/C (see filled symbols in Fig.~\ref{fig:model_params}). The thin lines  with downward arrows show the $1\sigma$ upper limits in the same configuration. For comparison, the purple dotted line shows the total relative uncertainties (errors combined quadratically) of AMS-02 F/Si ratio; see text for discussion.
   \label{fig:F_primfrac}}
\end{figure}
To understand whether this could be possible, it is interesting to look at the primary fraction of CR F (w.r.t. its total flux) in Fig~\ref{fig:F_primfrac}. The fraction grows steadily with the rigidity, which is expected as primary contributions grow $R^{\delta\approx 0.5}$ faster than secondary ones \citep[e.g.,][]{Vecchi:2022mpj}. All the curves are parallel because all our propagation configurations assume the same power-law index value for all primary species (and thus for the primary F). Were the F/Si data uncertainties constant with rigidity, the high-rigidity tip would be best to constrain the fluorine relative source abundance. However, because of decreasing fluxes, statistical uncertainties dominate at this end (see Fig.~\ref{fig:sub_errors}). For comparison purpose, we report the total uncertainties on AMS-02 F/Si (purple dotted line) in Fig~\ref{fig:F_primfrac}: we see that above 100~GV, the AMS-02 uncertainties are growing faster than the primary fraction; below this rigidity, the uncertainties are roughly constant, making the $\sim 10-100$~GV range the regime that constrains in practice the fluorine primary fraction.

Actually, because F is compatible with being a pure secondary (the fit only gives an upper limit), the primary fraction of F can be directly compared to the relative uncertainty on the F/Si data (that the model fits): the thin lines with downward arrows are roughly bounded by the dash-dotted purple line. The match is not exact and varies from cross-section model to another: this is because the minimisation procedure allows for a variation of the production cross sections; moreover the purple dotted line shows the total uncertainties (statistical and systematic uncertainties added in quadrature), whereas the analysis accounts for the covariance matrix of data uncertainties. In summary, to get tight enough constraints on the source abundance of F (to disentangle between the two source models), the goal of future experiments should be to reach---owing to the growing primary fraction with rigidity---either a $\lesssim 0.1\%$ precision on F/Si data at a few GV, or a few percent precision above tens of TV.

\section{Conclusions}
\label{sec:conclusions}

We have studied the AMS-02 F/Si data in a semi-analytical propagation model to constrain the transport parameters and the source abundance of $^{19}$F. We have updated the production cross sections of F from its main CR progenitors (Ne, Mg, Si, and Fe). We highlighted the importance of accounting for the anomalous isotopic abundance of $^{22}$Ne/$^{20}$Ne, in order not to bias the F production (impact $\lesssim 6\%$). For our analysis, we have propagated the uncertainties on several important input ingredients (Solar modulation and nuclear production cross sections), and have accounted for a best-guess covariance matrix of uncertainties for the F/Si data. The latter is dominated by a systematic with a significant correlation length (one rigidity decade) below a few ten of GV, which is equivalent to a global normalisation factor on these low-rigidity data. This directly impacts the statistical interpretation of the model, since using the usual quadratic distance estimator instead (i.e, no energy correlations in the data) would lead to worsened goodness-of-fit values.

We used several propagation setups (with or without convection and reacceleration, and two parametrisations of the low-rigidity break) and different production cross sections, but all configurations lead to the same conclusions. In a first step, we analysed AMS-02 F/Si data alone and found that (i) the F/Si data can be reproduced by the model assuming F is a pure secondary species (i.e. no astrophysical source of F); (ii) most of the transport parameters obtained from the F/Si analysis are consistent with those derived from the use of the traditional (Li,Be,B)/C ratios. In a second step, we have performed a combined analysis of F/Si and (Li,Be,B)/C), in order to break the (partial) degeneracy between the production cross sections and the normalisation of the diffusion coefficient. We found that (i) the global $\chi^2$ per degree of freedom was close to 1 and significantly better than that obtained from the (Li,Be,B)/C analysis only; (ii) the transport parameters were also slightly more constrained when adding F/Si data, as expected, with a slightly larger diffusion coefficient normalisation (but within the uncertainties); (iv) the posteriors obtained for the production cross-section parameters indicate a slight mismatch between the production of B and F, where a $10\%$ increase of B and  $5\%$ decrease of F are needed to best-fit the data. This is within the typical range of nuclear data uncertainties, and new or better nuclear data are needed to go further in this interpretation; in particular, F production data only have a single or pair of energy points for the whole energy domain, which is certainly not satisfactory. At this stage, we can nevertheless conclude that all AMS-02 secondary species with $Z\in[2,3,4,5,9]$ can be reproduced in a simple diffusion model.

Finally, we have obtained an upper limit on the relative source abundance of $^{19}$F. Fluorine is potentially a key element to test the underlying CR sources and their acceleration mechanisms, as it is one of the few light elements (with Na) for which the correlation between the volatility temperature and FIP breaks down; these two hypotheses are expected to lead to ${\rm (^{19}F/^{28}Si)}_{\rm CRS} \sim 2.2 \times 10^{-4}$ and $\sim 1.5 \times 10^{-4}$ respectively. Unfortunately, owing to the very small primary content in the CR flux of F ($\lesssim 5\%$ at 1~GV), we can only derive an upper limit at source of $\sim 5\times 10^{-3}$ from AMS-02 data; this limit is set by the data in the $50-100$~GV rigidity range (where the uncertainties are the smallest). In order to distinguish between the two above models, at least a factor $50$ improvement would be needed. This could be achieved by the challenging measurement of F/Si up to a few tens of TV at a few percent precision. This may be within reach of the next generation of CR experiments \citep{2019NIMPA.94462561S,2021ExA....51.1299B}.

\begin{acknowledgements}
We thank our CR colleagues at Annecy and Montpellier for discussions. This work was supported by the Programme National des Hautes Energies of CNRS/INSU with INP and IN2P3, co-funded by CEA and CNES.
We thank the Center for Information Technology of the University of Groningen for their support and for providing access to the Peregrine high-performance computing cluster.
\end{acknowledgements}

\bibliographystyle{aa} 
\bibliography{fluorine}

\clearpage
 \begin{appendix}

\section{$\chi^2$ analysis with covariance and nuisance}
\label{app:methodo}

In order to reduce biases in the model parameter determination, via a minimisation (Sect.~\ref{app:chi2_def}), three crucial ingredients are included \citep{2019A&A...627A.158D}: nuisance parameters for the Solar modulation level (Sect.~\ref{app:nui_phiff}) and for nuclear production cross sections (Sect.~\ref{app:nui_xs}), and the use of a covariance matrix of data uncertainties (Sect.~\ref{app:cov_mat}).

\subsection{$\chi^2$ minimisation}
\label{app:chi2_def}

Interstellar (IS) fluxes are obtained from the transport equation described in the previous section. To compare to the top-of-atmosphere (TOA) data, IS fluxes are Solar modulated. We use in this study the force-field approximation \citep{1967ApJ...149L.115G,1968ApJ...154.1011G}, in which
\begin{eqnarray}
  \psi^{\rm TOA}\left(E^{\rm TOA}\right) &=& \left(\frac{p^{\rm TOA}}{p^{\rm IS}}\right)^2 \psi^{\rm IS}\left(E^{\rm IS}\right)\,,\label{eq:psi_ff}
  \\
  E_{k/n}^{\rm TOA}&=&E_{k/n}^{\rm IS}-\frac{Z}{A}\phi_{\rm FF}\,.
  \label{eq:phi_ff}
\end{eqnarray}
The parameter $\phi_{\rm FF}$ in Eq.~(\ref{eq:phi_ff}) is the Fisk potential, which must be set appropriately for each data taking period (see below).

The best-fit model parameters and goodness-of-fit of the model to the data are obtained from a $\chi^2$ minimisation using \minuit~\citep{1975CoPhC..10..343J}, with
\begin{eqnarray}
  \chi^2 &=& \sum_{q=1}^{n_q} \left( {\cal D}^{\,q} + \sum_{s=1}^{n_s} {\cal N}^{s(q)}\right) + \sum_{x=1}^{n_x}{\cal N}^{\,x},
  \label{eq:chi2}\\
 {\cal D} &=& \sum_{i,j=1}^{n_E,n_E}\left(y^{\rm data}_i-y^{\rm model}_i\right) \left({\cal C}^{-1}\right)_{ij} \left(y^{\rm data}_j-y^{\rm model}_j\right),
     \label{eq:distance} \\
   {\cal N}(y) \!\!\!&=&\! \frac{\left(y-\mu_y\right)^2}{\sigma_y^2}.
   \label{eq:nuis}
\end{eqnarray}
In Eq.~(\ref{eq:chi2}), $q$ runs over the $n_q$ ratios used in the fit\footnote{For instance, $n_q=1$ for the stand-alone F/Si analysis (Sect.~\ref{sec:standaloneFSi}) and $n_q=4$ for the combined analysis of (Li,Be,B)/C and F/Si (Sect.~\ref{sec:B/C+F/Si}).}, for which the quadratic distance ${\cal D}$ between the model and the data, Eq.~(\ref{eq:distance}), is calculated including energy bin correlations ($n_E$ bins in total); these correlations are encoded in the covariance matrix ${\cal C}$ of data uncertainties discussed in Sect.~\ref{app:cov_mat}.
Gaussian-distributed nuisance parameters ${\cal N}$ of mean $\mu_y$ and variance $\sigma_y^2$, Eq.~(\ref{eq:nuis}), are considered on Solar modulation and some proxy cross sections (${\cal N}^s$ and ${\cal N}^x$ respectively): the indices $s$ and $x$ in  Eq.~(\ref{eq:chi2}) run over $n_s$ different data taking periods (if applies) and the $n_x$ cross-section reactions considered in the error budget respectively. The nuisance terms penalise the $\chi^2$ if the associated parameters wander several $\sigma$ away from their expected value.
Although we do not perform a Bayesian analysis, for brevity, we find useful to denote the nuisance parameters values $\mu_y$ and $\sigma_y^2$ priors and their post-fit values (found after the minimisation) posteriors.

The minimum $\chi^2$ value indicates how good is the fit: $\chi^2_{\rm min}/{\rm dof}\sim 1$ corresponds to an excellent fit, with $n_{\rm dof}=n_{\rm data}-n_{\rm pars}-n_s-n_x$.
As introduced in \citet{2020A&A...639A.131W}, it is also useful to consider
\begin{equation}
  \chipernui{} \equiv \left(\sum_{s=0}^{n_s}{\cal N}^s + \sum_{x=0}^{n_x}{\cal N}^{x}\right)/(n_s+n_x),
  \label{eq:chi2nuis}
\end{equation}
with $n_{\rm nui}=n_s+n_x$. The above quantity indicates whether the post-fit values stay very close to the priors ($\mu^{\rm post}\sim \mu^{\rm prior}$) or wander within $1\sigma$ of the priors ($\mu^{\rm post}\lesssim \mu^{\rm prior}\pm\sigma$), corresponding to $\chipernui{}\sim0$ or $\chipernui{}\lesssim 1$ respectively.

\subsection{Priors for Solar modulation}
\label{app:nui_phiff}

\begin{figure}[t]
  \includegraphics[width=\columnwidth]{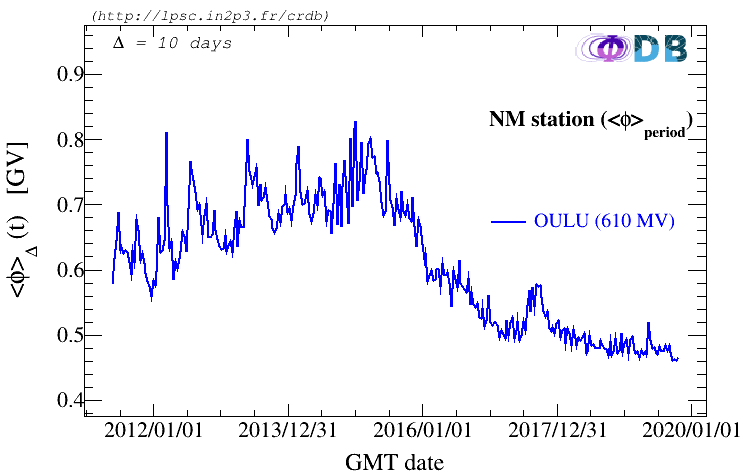}
  \caption{Solar modulation level reconstructed \citep{2017AdSpR..60..833G} from the OULU neutron monitor station, averaged over a 10 day period. The date range matches the period of AMS-02 analysed F/Si data \citep{2021PhRvL.126h1102A}, and the value in parenthesis (in the legend) gives the average modulation level over the period. Retrieved from \crdb{}.\label{fig:phi_series}}
\end{figure}

In the Force-Field approximation, Eqs~(\ref{eq:psi_ff}-\ref{eq:phi_ff}), modulation levels can be reconstructed from the analysis of neutron monitor (NM) data, at a precision $\sigma_{\phi}\simeq100$~MV \citep{2015AdSpR..55..363M,2017AdSpR..60..833G}. The reconstructed $\phi$ values are stored in the cosmic-ray database (\crdb{}\footnote{\url{https://lpsc.in2p3.fr/crdb}})  and can be accessed via the interface in the `Solar modulation' tab of \crdb{} \citep{2014A&A...569A..32M,2020Univ....6..102M}.  Figure~\ref{fig:phi_series} shows the modulation levels (averaged over 10 days) retrieved from \crdb{} from May 19, 2011 to October 30, 2019, for several for the OULU NM station.

The data considered in our analysis are either AMS-02 F/Si data only (see Sect.~\ref{sec:standaloneFSi}) or their combined analysis with AMS-02 Li/C, Be/C, and B/C data (see Sect.~\ref{sec:B/C+F/Si}). The data taking period of AMS-02 F/Si data \citep{2021PhRvL.126h1102A} matches that of Fig.~\ref{fig:phi_series}, i.e. 8,5 years of AMS-02 data, for which an average modulation level of $\mu_\phi=610$~MV is found (we take as reference the widely used OULU NM). For Li/C, Be/C, and B/C, the most recent published data  \citep{2021PhR...894....1A} are based on a 7 years data taking period ($\mu_\phi=636$~MV), but we nevertheless use the previously published data set based on 5 years of AMS-02 data \citep{2018PhRvL.120b1101A} for which $\mu_\phi=680$~MV. We do so because we wish to compare and combine the results of the F/Si analysis to that of our previous (Li,Be,B)/C analysis \citet{2020A&A...639A.131W,2022A&A...668A...7M}, which relies on the smaller dataset. This choice leads to slightly conservative error bars in our analyses, as AMS-02 datasets based on longer data taking periods have slightly smaller uncertainties.

We stress that, in principle, solar modulated fluxes should be calculated as weighted averages of fluxes modulated on short time intervals (for which the modulation level is roughly constant), whereas our calculation is based on an average modulation level over the full data taking period. However, the difference between the two calculations was shown to be at the few percent level on the low-energy fluxes, also amounting to a few MV difference on the average Solar modulation level used (see App.~A.2 of \citealp{2016A&A...591A..94G}). The latter number is much smaller than the uncertainty $\sigma_{\phi}\simeq100$~MV taken on the solar modulation level. We repeated the above comparison using 10-days slices over the period depicted in Fig.~\ref{fig:phi_series}. We reach similar conclusions for the fluxes and further find a negligible difference on ratios, at the level of $0.1\%$ in the AMS-02 rigidity range. In our analysis, we thus use for each dataset the NM-derived average modulation levels, and we recall that we apply the modulation on each CR isotope separately before forming the elemental ratios of interest.

\subsection{Priors for cross sections}
\label{app:nui_xs}

Production cross-section uncertainties are sizeable ($\sim5-20\%$), and the propagation equation involves a large network of reactions ($\gtrsim 1000$). Owing to the difficulty to assess the uncertainty on a reaction-to-reaction basis \citep{2018PhRvC..98c4611G}, and because different groups of reactions have similar impact on the quantity of interest (e.g. F/Si), a strategy to propagate the nuclear uncertainties in the calculation is to pick the most relevant reactions as proxies for the whole network \citep{2019A&A...627A.158D}. We first have to identify what these most important reactions are, and then to parameterise the uncertainties in a way that can be implemented as nuisance parameters.

In App.~\ref{app:rescaling}, we identify $^{20}$Ne, $^{24}$Mg, and $^{28}$Si as the most impacting progenitors. To turn the reactions into nuisance parameters, we follow the methodology detailed in \citet{2019A&A...627A.158D}, where transformation laws on cross sections combine a normalisation, energy scale, and low-energy slope, in order to modify a reference cross section $\sigma_{\rm ref}$:
\begin{flalign}
 \sigma^{\rm Norm.}(E_{k/n}) &= \mathrm{Norm} \times \sigma_{\rm ref}(E_{k/n})\,, \label{eq:NSS_Norm}\\
 \sigma^{\rm Scale}(E_{k/n}) &= \sigma_{\rm ref}\left(E_{k/n} \times \mathrm{Scale}\right)\,, \label{eq:NSS_Scale}\\
 \sigma^{\rm Slope}(E_{k/n}) &= \label{eq:NSS_Slope}
   \begin{cases}
    \displaystyle \sigma_{\rm ref} (E_{k/n})\times \left(\frac{E_{k/n}}{E_{k/n}^{\text{thr}}}\right)^\text{Slope}\!\!\! \text{if}\; E_{k/n} \leq E_{k/n}^{\text{thresh.}},\\
    \sigma_{\rm ref}(E_{k/n})\quad\quad \text{otherwise}\,.
  \end{cases}
\end{flalign}
In the above equations, $E_{k/n}^{\text{thresh.}}$ is fixed to 1~GeV/n. The parameters `Norm', `Scale' and `Slope' (NSS) are the ones for which we need to determine a central value $\mu$ and dispersion $\sigma$, to be used as nuisance parameters in Eq.~(\ref{eq:chi2}).

\begin{figure}[t]
   \includegraphics[scale=0.395]{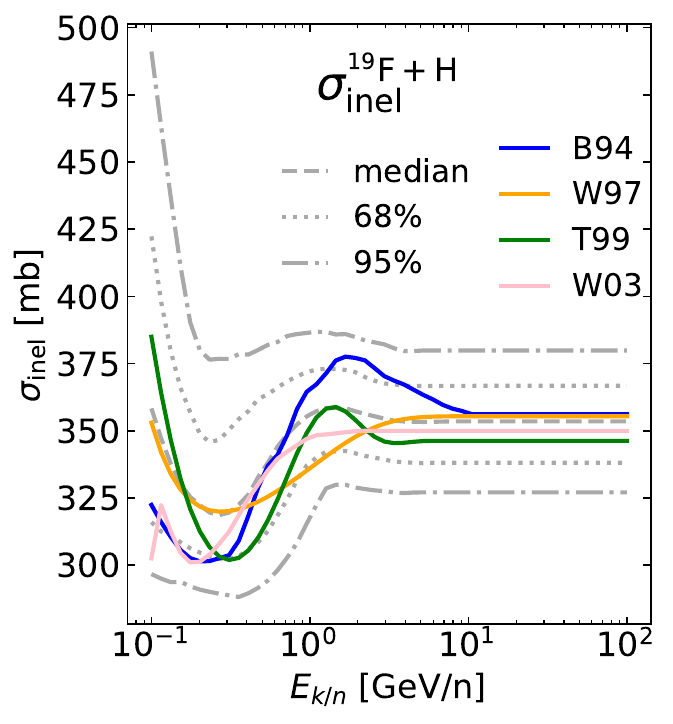}
   \includegraphics[scale=0.395]{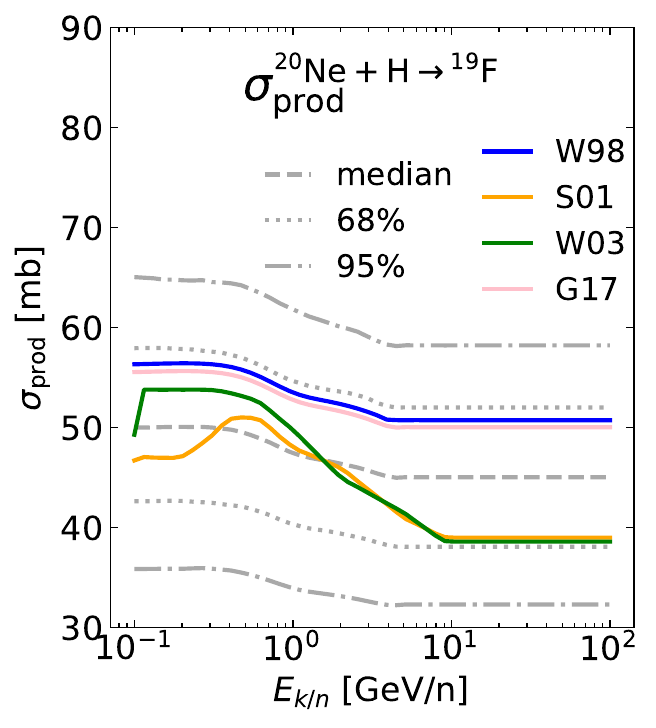}
   \caption{Illustration of the NSS scheme used for cross section nuisance parameters (inelastic on the left and production on the right), see Eqs.~(\ref{eq:NSS_Norm}-\ref{eq:NSS_Slope}). Colour-coded solid lines correspond to existing cross section parametrisations: inelastic on the left, {\tt B94} \citep{BarPol1994}, {\tt W97} \citep{1996PhRvC..54.1329W}, {\tt T99} \citep{1997lrc..reptQ....T,1999STIN...0004259T}, and {\tt W03}~\citep{2003ApJS..144..153W}; production on the right, \xsWa{} \citep{1998ApJ...508..940W,1998ApJ...508..949W,1998PhRvC..58.3539W}, \xsS{} (A. Soutoul, private communication), \xsW{}~\citep{2003ApJS..144..153W}, and {\tt G17}~\citep{2001ICRC....5.1836M,2003ICRC....4.1969M}. Grey lines correspond to the median, $68\%$, and $95\%$ CLs resulting from varying the NSS parameters: we choose for $\sigma_{\rm NSS}$ of the cross-section nuisance parameters (reported in Table~\ref{tab:xs_nuis}), the NSS values leading to the 68\% CLs (dotted grey lines).}
   \label{fig:xs_nuis}
\end{figure}
Because we renormalise production cross-section reactions to nuclear data in App.~\ref{app:rescaling}, we enforce $\mu_{\rm Norm}=1$, $\mu_{\rm Scale}=1$, and $\mu_{\rm Slope}=0$ for all reactions.
To fix $\sigma$, as illustrated in Fig.~\ref{fig:xs_nuis}, we visually estimate what transformation law values are needed to encompass at $1\sigma$ (dotted grey lines) the different cross-section parametrisations (coloured lines)\footnote{This crude approach possibly overestimates the uncertainty seen in nuclear data. However, we recall that we use only a few reactions as proxies of all reactions involved in the calculation. Moreover, very few nuclear data points are available, even for the most important reactions (see App.~\ref{app:rescaling}). All in all, $\sigma$ is difficult to evaluate and anyway, varying its value within reason would not change the conclusions of our analyses.}.
\begin{table}[t]
\caption{Reaction proxies and values of their nuisance parameters.}
\label{tab:xs_nuis}
\centering
{
\footnotesize
\begin{tabular}{l c c c c c c}
\hline\hline
Reaction &\hspace{-0.75cm}Impact $(\frac{\Delta\sigma}{\sigma})^{\rm XS}$\hspace{-0.75cm}  & Norm. & Scale & Slope \\
         & on F/Si & $\; \mu \,|\, \sigma$ & $\mu \,|\, \sigma$ & $\mu \,|\, \sigma$\\
\hline
\boldmath{$^{19}$}{\bf F+H}  & {\bf 4\%} & 1.0\,|\,0.04 & 1.0\,|\,0.5 &  n/a \\
$^{20}$Ne+H & 2\%       & 1.0\,|\,0.06 & 1.0\,|\,0.5 &  n/a \\
$^{24}$Mg+H & 1\%       & 1.0\,|\,0.07 & 1.0\,|\,0.4 &  n/a \\
$^{28}$Si+H & 4\%       & 1.0\,|\,0.07 & 1.0\,|\,0.4 &  n/a \\
All$^{(a)}$   & 6\%       & \multicolumn{3}{c}{} \\[3mm]
\boldmath{$^{20}$}{\bf Ne\,+H}\boldmath{$\veryshortarrow$$^{19}$}{\bf F [30\%]}  & {\bf 10\%} & 1.0\,|\,0.13 &  n/a  & 0.0\,|\,0.01 \\
$^{24}$Mg+H$\veryshortarrow$$^{19}$F [15\%]    & 5\%        & 1.0\,|\,0.33 &  n/a  & 0.0\,|\,0.04 \\
$^{28}$Si\;+\;H$\veryshortarrow$$^{19}$F [15\%]& 3\%        & 1.0\,|\,0.27 &  n/a  & 0.0\,|\,0.10 \\
All$^{(a)}$   & 20\%   & \multicolumn{3}{c}{} \\
\hline
\end{tabular}
}
\tablefoot{
For the most relevant inelastic (upper half) and production (lower half) reactions, we show the typical impact nuclear uncertainties have on the calculated F/Si ratio (col.~2), and the nuisance parameters $\mu$ and $\sigma$ for the NSS transformation laws (cols~3-5), see Eqs~(\ref{eq:NSS_Norm}-\ref{eq:NSS_Slope}). For the production reactions, we also show their contributing fraction w.r.t the total F production (in square brackets, col.~1, as read off Fig.~\ref{fig:xs_prodchannels}). In practice, for the F/Si $\chi^2$ analysis, only two reactions (highlighted in {\bf boldface}) are chosen as proxies.\\
\tablefoottext{a}{Impact of changing cross sections for all reactions at once for the F/Si calculation.}
}
\end{table}
Table~\ref{tab:xs_nuis} gathers the NSS $\mu$ and $\sigma$ values to use for the production and inelastic reactions listed; $\sigma$ values for Li, Be, and B production can be found in Table~B.1 of \citet{2020A&A...639A.131W}.
The second column also highlights the typical uncertainties on F/Si observed when using different modelling for the reactions listed in the first column.

We checked (not shown) that the various production reactions (and inelastic reactions among themselves) have similar impact on the F/Si shape and mostly differ on the amplitude of the change. This leads to quasi-degenerate parameters for the many cross-section reactions, i.e. the same production rate can be obtained by increasing one reaction cross section and decreasing another \citep{2019A&A...627A.158D}. As strong degeneracies are difficult to tackle by minimisation algorithms, the full list of reactions identified in Table~\ref{tab:xs_nuis} is not used in practice (and furthermore, reactions whose impact is smaller than the data uncertainties are discarded). Consequently, we use a single proxy for the inelastic cross sections ($^{19}$F+H) and the production cross sections ($^{20}$Ne\,+H$\veryshortarrow$$^{19}$F). These proxies are highlighted in boldface in Table~\ref{tab:xs_nuis}, and compared to the row denoted `All' that shows the impact of the uncertainties from the full network.

\subsection{Covariance matrix of uncertainties}
\label{app:cov_mat}

An important term in Eq.~(\ref{eq:chi2}) is the covariance matrix of data uncertainties $C_{ij}$. Because the AMS collaboration does not provide such a matrix, almost all analyses of their data rely on total uncertainties. However, as shown on simulated data in \citet{2019A&A...627A.158D}, not accounting for the covariance matrix can significantly bias the determination of the transport parameters; the crucial role of the covariance matrix was further demonstrated in the context of analysing AMS-02 antiprotons data \citep{2020PhRvR...2b3022B,2020PhRvR...2d3017H}.

To remedy this situation, we can build a tentative covariance matrix of uncertainties based on the information provided in the AMS-02 publications and supplemental material. Following \citet{2019A&A...627A.158D}, we define the relative covariance $(C_{\rm rel}^\alpha)_{ij}$ between rigidity bin $R_i$ and $R_j$ (and the associated correlation matrix) to be
\begin{eqnarray}
(C_{\rm rel}^\alpha)_{ij} &=& \sigma^\alpha_i \sigma^\alpha_j \exp\left(-\frac{1}{2}
\frac{(\log (R_i/R_j)^2} {(\ell_\alpha)^2} \right)\,,
\label{eq:cov}\\
{\rm c}_{ij}^{\alpha} &\equiv& \frac{{\cal C}_{ij}^{\alpha}}{\sqrt{{\cal C}_{ii}^{\alpha} \times {\cal C}_{jj}^{\alpha}}}\,,
\label{eq:correl}
\end{eqnarray}
with $\sigma_i^\alpha$ the relative uncertainty of error type $\alpha$ at bin $i$ and $\ell_\alpha$ the correlation lengths for error type $\alpha$ (in unit of rigidity decade).

The AMS-02 collaboration provides the statistical and systematic uncertainties split in three different components (acceptance, unfolding\footnote{Unfolding is the procedure to estimate `true' rigidities from measured ones, owing to the finite energy resolution of the detector.}, and scale), as displayed in solid lines in the top panel of Fig.~\ref{fig:sub_errors}. With the detailed information given in the supplemental material \citep{2021PhRvL.126h1102A}, we can estimate the correlation length associated with these systematic; see \citet{2019A&A...627A.158D} for more details. We set $\ell_{\rm scale}=\infty$ for the energy scale uncertainty, amounting to a global rigidity shift, and $\ell_{\rm Unfolding}=0.5$, a mild correlation resulting from the unfolding procedure.
The acceptance uncertainty combines errors of different origins, which motivates its decomposition in three extra components \citep{2019A&A...627A.158D}: (i) `Acc. norm.' is associated to inelastic cross-section uncertainties (impacting the detector acceptance estimation), with a relatively large correlation length $\ell_{\rm Acc.~norm.}=1$; (ii) `Acc. LE' is a low-rigidity error associated to the orbit-varying rigidity cut-off of the AMS detector, set to $\ell_{\rm Acc.~LE}=0.3$; and (iii) `Acc. res.' is a residual error mostly associated with data/Monte Carlo corrections. The latter cannot be associated to a particular physics process and no prescription can be taken for its correlation length.

As can be seen in the top panel of Fig.~\ref{fig:sub_errors}, the `Acc. norm.' (dashed line) and `Acc. res.' (dash-dotted line) uncertainties dominate the error budget at intermediate rigidities; a quite similar ranking is seen in (Li,Be,B)/C data, see Fig.~A.1 in \citet{2020A&A...639A.131W}. As studied on the AMS-02 B/C data in \citet{2019A&A...627A.158D}, taking extreme values of the unknown $\ell_{\rm Acc.~res.}$ impacts the best-fit \chimindof{} values (this value increases with $\ell_{\rm Acc.~res.}$), but only marginally the best-fit parameters and thus the best-fit flux ratios. For this reason, the \chimindof{} values we obtain (see Sect.~\ref{sec:B/C+F/Si}) and their interpretation  must be taken with a grain of salt. Otherwise, our conclusions are expected to be independent of the exact value of this correlation length, and for definiteness, we set $\ell_{\rm Acc.~res.}=0.1$ \citep{2019A&A...627A.158D}. The bottom panel of Fig.~\ref{fig:sub_errors} shows the corresponding covariance matrix of uncertainties accounting for all uncertainties: below 10~GV, the data are typically correlated over one decade in energy (dominated by `Acc. norm.'); above, we transition from weakly correlated (dominated by `Acc.~res.' to uncorrelated as statistical uncertainties take over.

\begin{figure}[t]
\centering
  \includegraphics[width=\columnwidth]{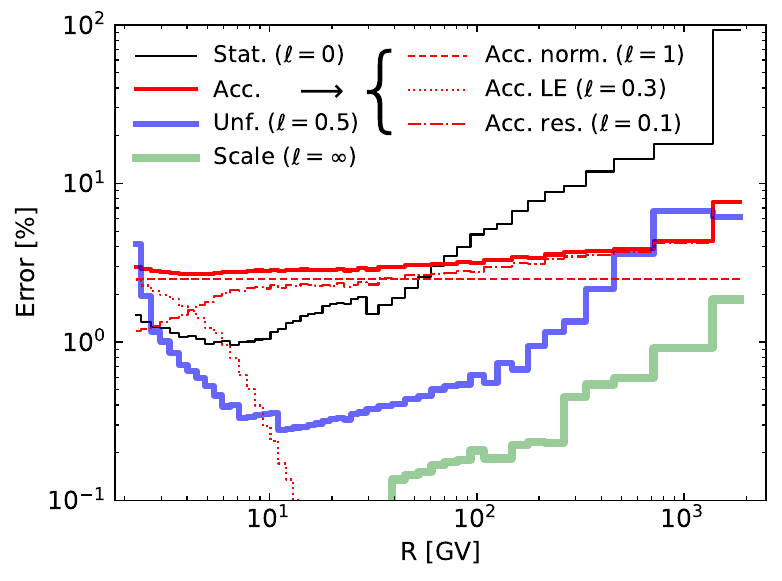}
  \includegraphics[width=0.9\columnwidth]{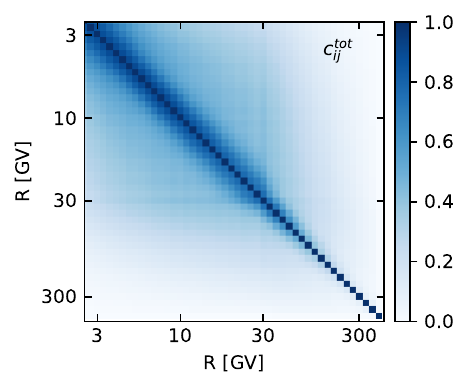}
  \caption{{\bf Top panel}: statistical and Acc., Unf., and Scale systematic uncertainties for F/Si, taken from Table S1 of \citet{2021PhRvL.126h1102A}. The `Acc. norm.' (dashed line), `Acc. LE' (dotted line), and `Acc. res.' (dash-dotted line) systematics are broken down from the `Acc.' systematic as explained in the text.  {\bf Bottom panel:} correlation matrix ${\rm c}_{ij}^{\rm tot}$, Eq.~(\ref{eq:correl}), for the combined statistical and systematics uncertainties, colour-coded from no correlation (white, ${\rm c}_{ij}^{\rm tot}=0$) to full correlations (blue, ${\rm c}_{ij}^{\rm tot}=1$).
  \label{fig:sub_errors}}
\end{figure}


\section{Production cross sections and F progenitors}
\label{app:prod_xs}
In this Appendix, we focus on the production cross sections. We show first the nuclear parametrisations renormalised on nuclear data points for a selection of important reactions (App.~\ref{app:rescaling}). We then illustrate the importance of correctly setting the source of Ne isotopes for the F/Si calculation (App.~\ref{app:anomaly}). With these updated cross sections and correct CR source settings, we determine the ranking of the most important progenitors of F (App.~\ref{app:ranking}).

\subsection{Rescaling of cross-sections on data}
\label{app:rescaling}

Several production cross-section parametrisations are available from the literature. In particular, the widely used \galprop{}\footnote{\url{https://galprop.stanford.edu/}} dataset \citep{2001ICRC....5.1836M,2003ICRC....4.1969M} is based on systematic fits on existing nuclear data of parametric models: the underlying models are either the semi-empirical formulae implemented in the \wnew{} code \citep{1998ApJ...508..940W,1998ApJ...508..949W,1998PhRvC..58.3539W} or the semi-analytical formulae implemented in the \yieldx{} code \citep{1998ApJ...501..911S,1998ApJ...501..920T}; these models lead to the \xsGalxii{} and \xsGalxxii{} production datasets respectively.
These parametrisations were established almost two decades ago, while several nuclear datasets became available since. Indeed, as shown in \citet{2022A&A...668A...7M} in the context of Li, Be, and B production, some reactions in these models are at odds with these new nuclear data. In such case, the natural procedure is to rescale these reactions to the new data.
We apply this normalisation procedure on $^{56}{\rm Fe}$, $^{32}{\rm S}$, $^{28}{\rm Si}$, $^{27}{\rm Al}$, $^{24}{\rm Mg}$, and $^{20,22}{\rm Ne}$ interacting on H (the most important progenitors of F, see App.~\ref{app:ranking}), to give either directly $^{19}$F (the only stable isotope of Fluorine) or short-lived nuclei ending up their decay chain\footnote{While we work in this section on the individual production cross sections $\sigma$, we stress that, in the rest of the paper, cumulative cross sections $\sigma^c$ are considered instead (relevant ones for CR propagation), with $\sigma^{\rm c}(X\to Y) = \sigma(X\to Y) +\sum_G\sigma(X\to G)\cdot {\cal B}r\,(G\to Y)$, where $G$ runs over the list of short-lived nuclei for $Y$ with a branching ratio ${\cal B}r$.} into $^{19}$F: the relevant nuclei with ${\cal B}r>5\%$ are \citep{1984ApJS...56..369L,00/04/78/51/PDF/tel-00008773.pdf} $^{19}{\rm Ne}$  (${\cal B}{\rm r}=100\%$), $^{19}{\rm O}$  (${\cal B}{\rm r}=100\%$), $^{19}$C (${\cal B}{\rm r}=20.9\%$), $^{20}$C (${\cal B}{\rm r}=48.6\%$), $^{19}$N (${\cal B}{\rm r}=45.4\%$), and $^{20}$N (${\cal B}{\rm r}=57\%$).

\begin{figure}[t]
   \centering
   \includegraphics[trim={0  44 6 3},clip,width=0.549\columnwidth]{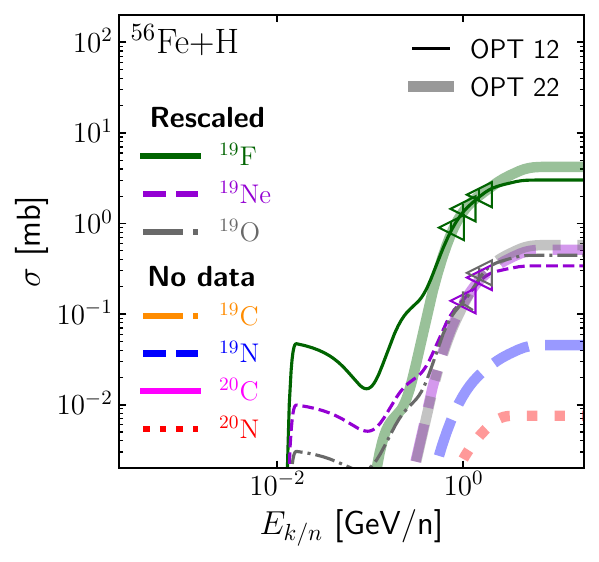}
   \hspace{1.cm}
   \includegraphics[trim={0 -40 -30  0},clip,width=0.3\columnwidth]{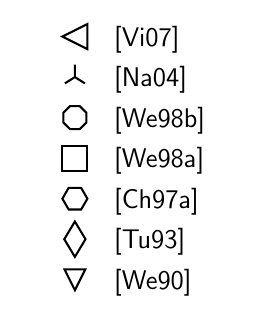}\\
   \includegraphics[trim={0  44 6 3},clip,width=0.549\columnwidth]{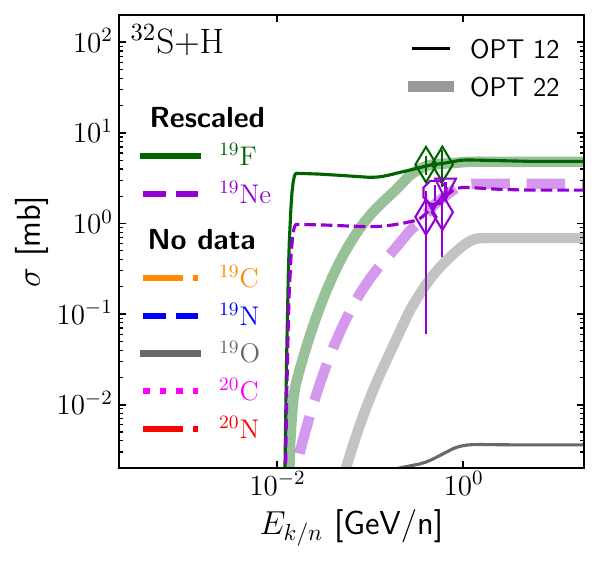}
   \includegraphics[trim={55 44 6 3},clip,width=0.440\columnwidth]{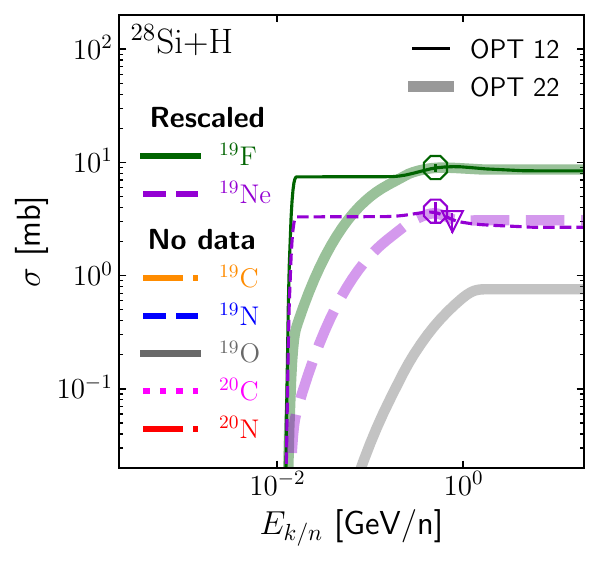}\\
   \includegraphics[trim={0  44 6 3},clip,width=0.549\columnwidth]{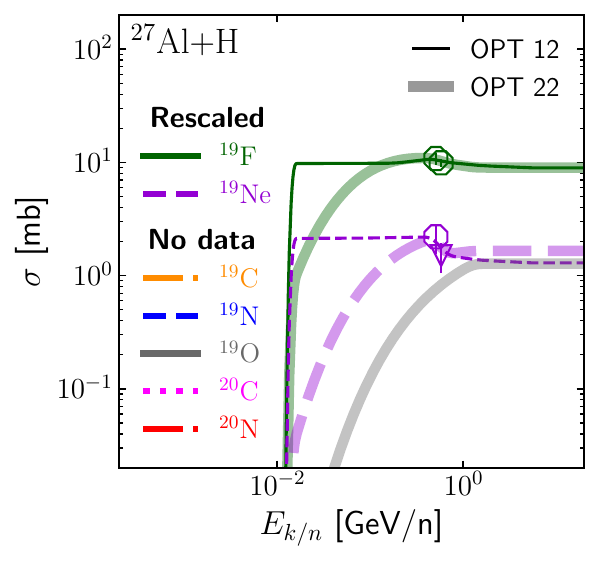}
   \includegraphics[trim={55 44 6 3},clip,width=0.440\columnwidth]{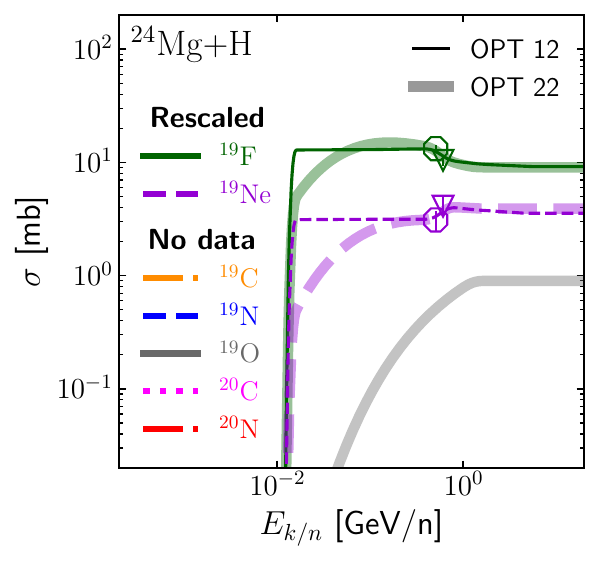}\\
   \includegraphics[trim={0  8 6 3},clip,width=0.549\columnwidth]{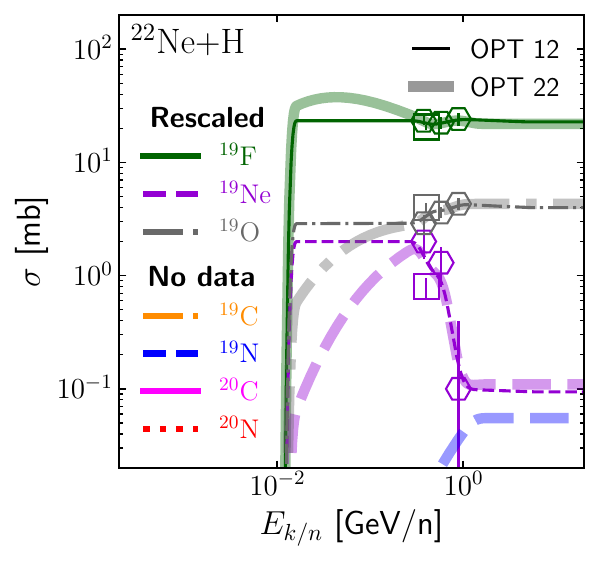}
   \includegraphics[trim={55 8 6 3},clip,width=0.440\columnwidth]{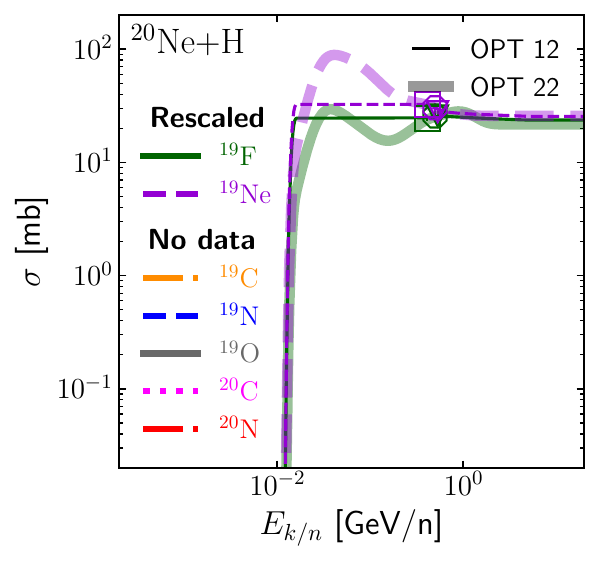}
   \hspace{2cm}
   \caption{Models (\optxii{} and \optxxii{}) and data (symbols) for the colour-coded production of $^{19}$F  and ghosts with ${\cal B}r>5\%$ (i.e. $^{20}$N, $^{20}$C, $^{19}$Ne, $^{19}$O, $^{19}$N, $^{19}$C), from $^{56}$Fe, $^{32}$S $^{28}$Si, $^{27}$Al, $^{24}$Mg, and $^{20}$Ne CRs on H.
   `No data' indicates that no nuclear data were found for the production of the isotopes listed, hence the model corresponds to the original parametrisation (if not visible on the plot, it means that the cross-section value is $\lesssim10^{-2}$~mb). `Rescaled' indicates that the isotopes listed have nuclear data on which the model were renormalised. For the latter, we do not show the original model values for these reactions (for readability), but refer the reader to Table~\ref{Tab:F_XS_rescalingfactor_HE} that highlights the rescaling factor between the new and the original cross sections in the asymptotic regime.
   The data references (top right panel) correspond to \citet{2007PhRvC..75d4603V}, \citet{2004PhRvC..70e4607N}, \citet{1998ApJ...508..949W,1998PhRvC..58.3539W}, \citet{1997ApJ...479..504C}, \citet{1993ICRC....2..163T}, and \citet{1990PhRvC..41..547W}.
   \label{fig:xs_data_vs_model}
   }
\end{figure}

In practice, for this nuclear cross-section renormalisation, we (i) start from the \optxii{}, \optxiiupxxii{}, and \optxxii{} datasets (updated for Li, Be, and B production) provided in \citet{2022A&A...668A...7M}, (ii) extract from the \exfor{} database\footnote{\url{https://www.nndc.bnl.gov/exfor/exfor.htm}} \citep{2014NDS...120..272O} all nuclear data relevant for the F production (on H targets), (iii) apply the rescaling procedure of \citet{2022A&A...668A...7M} to get updated modelling of the above reaction cross sections, (iv) use the empirical formulae of \citet{1988PhRvC..37.1490F} for $\sigma^{\rm He}/\sigma^{\rm H}$ to get the corresponding cross sections on He, and (v) generate production cross-section sets for the cumulative reactions, to be used in Sect.~\ref{sec:B/C+F/Si}.

The nuclear data along with the rescaled \optxii{} and \optxxii{} energy-dependent parametrisations are shown in 
Fig.~\ref{fig:xs_data_vs_model} (thin and thick lines respectively), along with the data (symbols). As can be seen, for most reactions, only one or two nuclear data points (at similar energies) are available. In that case, the updated cross sections amount to a mere rescaling of the original ones---the rescaling values are gathered in Table~\ref{Tab:F_XS_rescalingfactor_HE} and discussed below. Our procedure significantly changes the original energy dependence of the cross section for $^{22}$Ne+H$\rightarrow^{19}$Ne only (dashed purple line in the bottom left panel): the asymptotic high-energy value of this reaction is quite uncertain, as it is fixed by the highest-energy nuclear data point (which suffers large uncertainties); this reaction would particularly benefit from new data.
For the selection of the most important progenitors of F, we first remark that: (i) very few data exist and on a very limited energy range; (ii) although they are expected to play a mostly negligible part in the overall production of F, most of the associated ghost nuclei ($^{19}$C, $^{20}$C, $^{19}$N, and $^{20}$N) have no data at all; (iii) the production into $^{19}$O (ghost nucleus) contributes to $\sim10\%$ of the cumulative cross sections into F, but nuclear data are missing for half of the progenitors shown. Concerning the models, the parametrisations \optxii{} (thin lines) and \optxxii{} (thick lines) differ at low-energy. This is below the energy range of AMS-02 F data, but because the renormalisation point is in the rising regime, the high-energy asymptotic value (relevant for AMS-02 data) can be significantly different. We stress that \optxxii{} has a more realistic behaviour near the production threshold, so it should be favoured. Finally, for unmeasured reactions, the amplitude of these two parametrisations can also differ a lot (e.g. $^{27}$Al into $^{19}$F, compare the thin and thick blue dash-dotted lines).
\begin{table}
\centering
\caption{Rescaling factors applied on the main reactions producing F.}
\label{Tab:F_XS_rescalingfactor_HE}
\begin{tabular}{lccc}
\hline\hline
                & $^{19}{\rm F}$ & $^{19}{\rm Ne}$ & $^{19}{\rm O}$ \\
                &                &   (${\cal B}{\rm r}=100\%$)     &   (${\cal B}{\rm r}=100\%$)    \\
\hline
$^{56}{\rm Fe}$ &    5.2|0.6     &    1.92|0.50    &    130|0.7     \\
$^{32}{\rm S}$  &    0.6|0.6     &    1.04|1.03    &    $\times$    \\
$^{28}{\rm Si}$ &       1        &    0.91|0.90    &    $\times$    \\
$^{27}{\rm Al}$ &     0.98       &       0.87      &    $\times$    \\
$^{24}{\rm Mg}$ &   0.88|0.90    &       1.19      &    $\times$    \\
$^{22}{\rm Ne}$ &   1.16|1.45    &       0.1       &    1.18|1.15   \\
$^{20}{\rm Ne}$ &   1.03|0.98    &    0.94|0.95    &    $\times$    \\
\hline
\end{tabular}
\tablefoot{
  Rescaling factors ${\cal R}_{\rm model}=\sigma^{\rm new}_{\rm model}/\sigma^{\rm old}_{\rm model}$ at high energy for the production of F (the index {\em model} is set to either \optxii{} or \optxxii{}). The first column shows the list of progenitors (or projectiles) while the first line shows the list of the main fragments (stable isotope and ghosts with ${\cal B}{\rm r}=100\%$); other significant ghosts of F with ${\cal B}{\rm r}>5\%$, $^{19}$C (${\cal B}{\rm r}=20.9\%$), $^{20}$C (${\cal B}{\rm r}=48.6\%$), $^{19}$N (${\cal B}{\rm r}=45.4\%$), and $^{20}$N (${\cal B}{\rm r}=57\%$) are not shown, because no nuclear data are available for any of the projectiles considered. Entries in columns 2-4 show two factors, one for each {\em model}, or only one value if $\sigma^{\rm old}_{\tt OPT12}=\sigma^{\rm old}_{\tt OPT22}$. The key $\times$ means no-data available (hence $\sigma^{\rm new}=\sigma^{\rm old}$).
}
\end{table}

To better highlight the difference of our new production sets compared to the original ones, we show in Table~\ref{Tab:F_XS_rescalingfactor_HE} their ratios above a few GeV/n, in a regime where cross sections are assumed to be constant. For most progenitors, the new cross sections are within a few percent below or above the old ones, which was expected (i.e. no change w.r.t. the original \galprop{} cross sections). However, there are three important changes: first, it appears that the production of $^{19}$F from $^{56}$Fe was significantly underestimated in \optxii{} and overestimated in \optxxii{} (left and right pipe-separated numbers in the Table); second, the production  of $^{19}$F from $^{32}$S was also overestimated; third, the production of the $^{19}$Ne ghost from $^{22}$Ne is decreased by a factor ten\footnote{Looking at the bottom left panel of Fig.~\ref{fig:xs_data_vs_model}, we see that this originates from a single point (orange hexagon on top of the dashed lines) with very large error bars.}.

As obvious from the plots and the Table, gathering new nuclear data for the production of $^{19}$F would be a huge improvement to interpret the high-precision F/Si (and F) AMS-02 data.

\subsection{Impact of the anomalous $^{22}$Ne/$^{20}$Ne ratio on F/Si}
\label{app:anomaly}

In all our previous studies dealing with light nuclei, namely (Li,Be,B)/C \citep{2010A&A...516A..66P,2015A&A...580A...9G,2017PhRvL.119x1101G,2019PhRvD..99l3028G,2020A&A...639A.131W,2001ApJ...555..585M,2010A&A...516A..67M,2022A&A...668A...7M}, elemental abundances were rescaled to match the corresponding elemental CR fluxes, though keeping in the process isotopic abundances fixed to their Solar system (SS) fractions \citep{2003ApJ...591.1220L}. This is no longer possible here, because Ne, one of the most important progenitors for F (see next section), has an anomalous isotopic source abundance. Indeed, a detailed analysis shows that $(^{22}{\rm Ne}/^{20}{\rm Ne})_{\rm CRS} / (^{22}{\rm Ne}/^{20}{\rm Ne})_{\rm SS} \approx 4$ \citep{2005ApJ...634..351B}, indicating a different nucleosynthetic origin for $^{22}$Ne---possibly from Wolf-Rayet stars in OB associations. Other isotopic anomalies at source have been spotted, such as $^{12}$C/$^{16}$O and $^{58}$Fe/$^{56}$Fe \citep{1981PhRvL..46..682W,2005ApJ...634..351B,2008NewAR..52..427B}, but they are smaller than the Ne anomaly and furthermore not relevant for the F/Si ratio. In passing, we recall that being able to characterise an anomaly depends on the confidence we have on the production cross section (see for instance \citealt{1997ApJ...479..504C} for an illustration) and to a lesser extent, to our capability in deconvolving acceleration segregation mechanisms when comparing the propagated source abundances to CR data \citep{1997ApJ...487..182M,2021MNRAS.508.1321T}.

\begin{figure}[t]
  \includegraphics[width=\columnwidth]{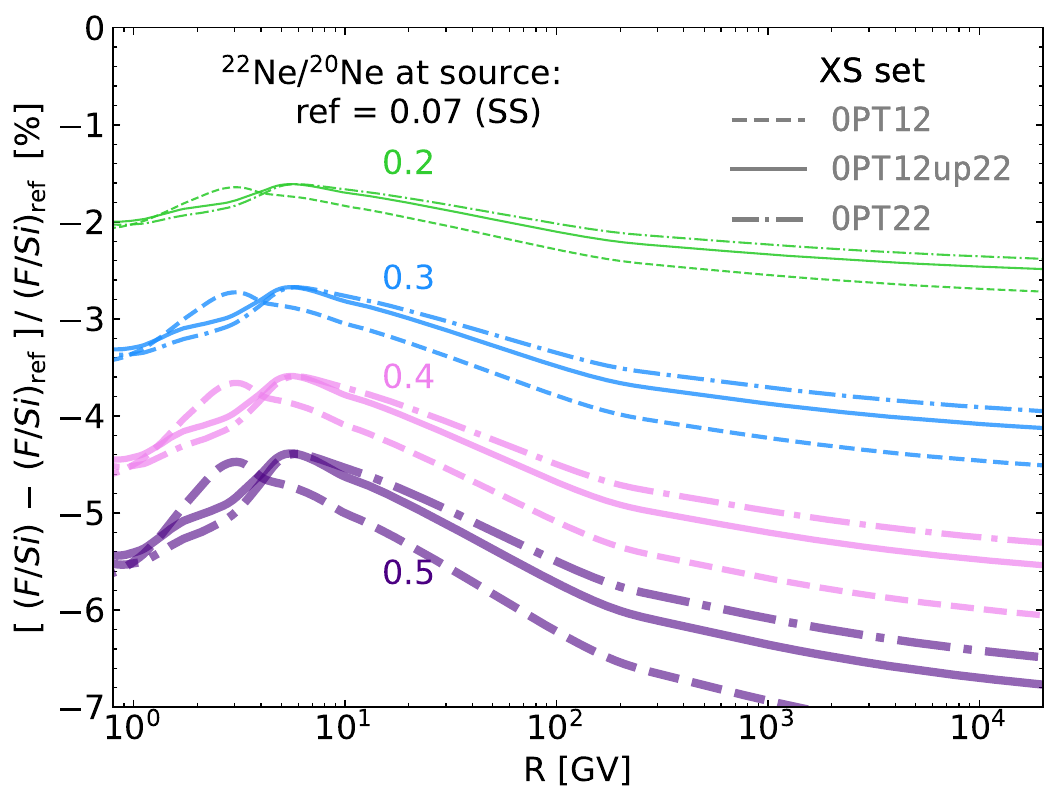}
  \caption{Impact of changing the isotopic ratio $(^{22}{\rm Ne}/^{20}{\rm Ne})_{\rm CRS}$ at source on the propagated F/Si ratio, illustrated for three equally plausible updates of the F production cross sections (different line styles). The reference calculation is based on the Solar system value \citep{2003ApJ...591.1220L}, and the various curves (from top to bottom) illustrate that growing values lead to decreasing F/Si ratios. See text for discussion.
  \label{fig:impact_22to20Ne}}
\end{figure}
Figure~\ref{fig:impact_22to20Ne} illustrates the importance of using the correct isotopic abundance for the F/Si calculation. With respect to the calculation based on the Solar System isotopic abundance $(^{22}{\rm Ne}/^{20}{\rm Ne})_{\rm SS}\simeq0.07$, the F/Si ratio decreases when the relative fraction of $^{22}$Ne to $^{20}$Ne at source grows. This is understood from the inspection of the bottom plots of Fig.~\ref{fig:xs_data_vs_model} where, for the cumulative production of F, we have $\sigma^c(\mathrm{^{22}Ne+H\to^{19}F})=26$~mb and $\sigma^c(\mathrm{^{20}Ne+H\to^{19}F})=47$~mb.
The typical $3\%$ to $6\%$ difference in F/Si, compared to the case of using the SS value is significant in the context of F/Si AMS-02 data, for which the best precision is $\sim 3\%$ (see the top panel of Fig.~\ref{fig:sub_errors}).

It is interesting to compare the $\rm(^{22}Ne/^{20}Ne)_{\rm CRS}$ anomaly obtained from different authors, based on the same ACE-CRIS Ne isotopic data. In the original analysis of these data, \citet{2005ApJ...634..351B} used a leaky-box model with \citet{1998ApJ...501..911S}'s production cross sections rescaled to existing nuclear data (that would be similar to our \optxxii{} set, minus the use of recent nuclear data). They obtained $\rm(^{22}Ne/^{20}Ne)_{\rm CRS}=0.39$, to compare to 0.32 reported by the \galprop{} team using \xsGalxii{} production set \citep{2020ApJS..250...27B}. At variance, we obtain here $\rm(^{22}Ne/^{20}Ne)_{\rm CRS}=0.47$  (see Sect.~\ref{sec:Fsource}). It is beyond the scope of this paper to investigate further the origin of this difference, which could possibly be related to the updated production cross sections.

\subsection{Ranking the most important progenitors}
\label{app:ranking}

Equipped with the updated production cross sections and using the appropriate rescaling for the isotopic source abundances of Ne, we can now precisely determine the most important progenitors of the CR flux of F, assuming all F is secondary in origin.
To do so, we follow the methodology proposed in \citet{2018PhRvC..98c4611G} and \citet{2022A&A...668A...7M}.

\begin{figure}[t]
  \includegraphics[width=\columnwidth]{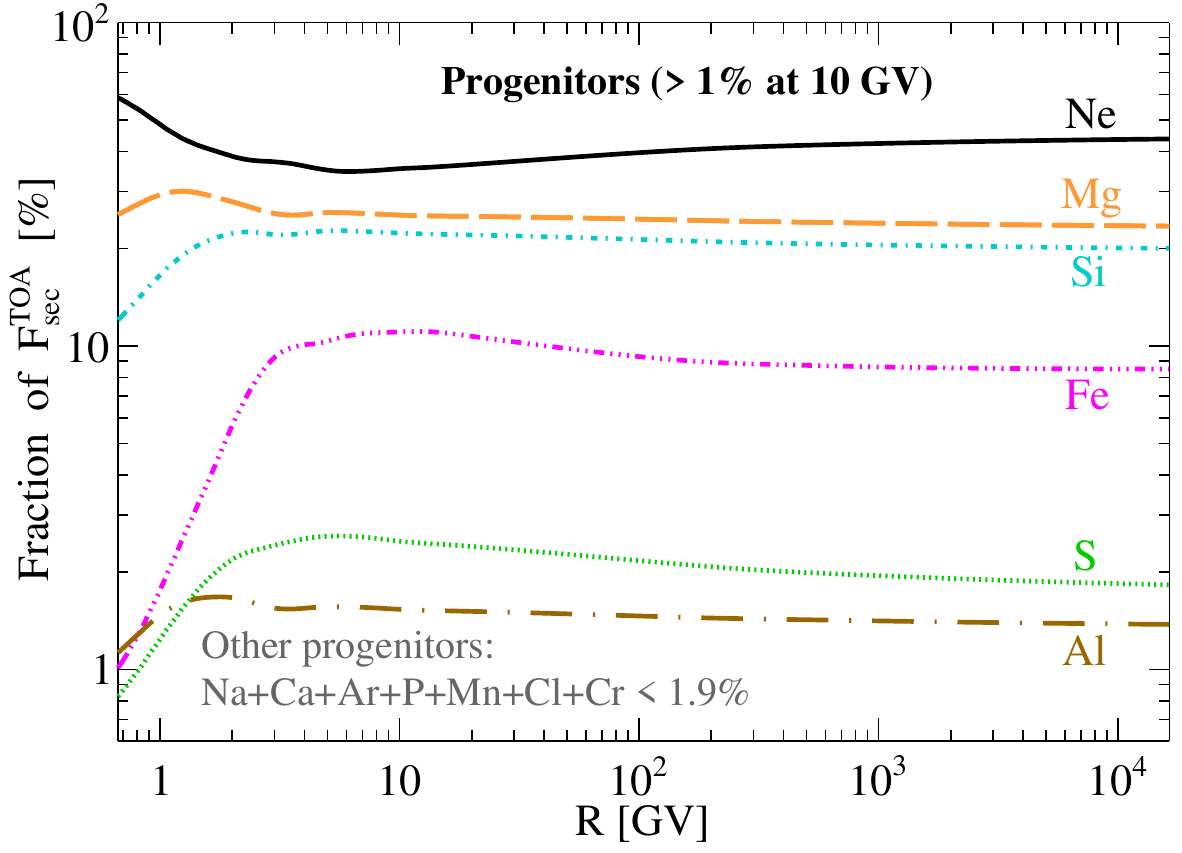}
  \includegraphics[width=\columnwidth]{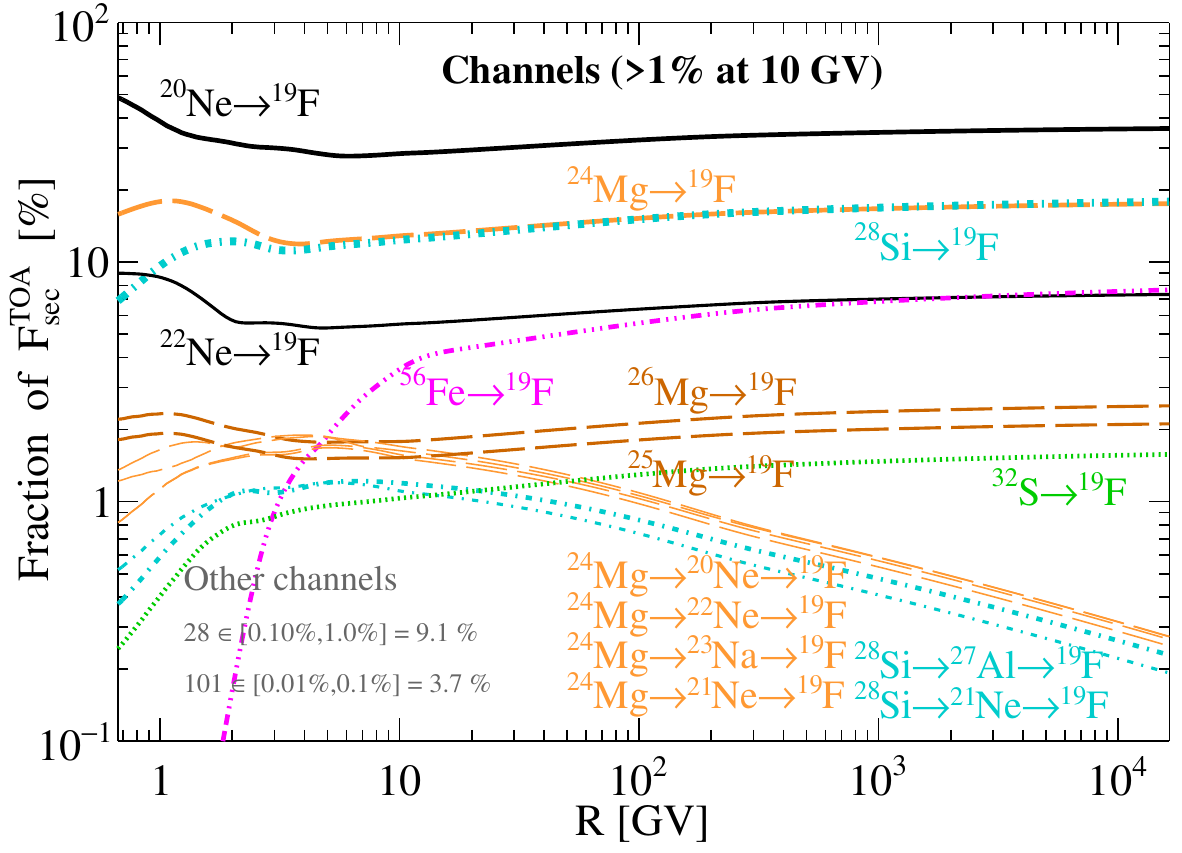}
  \caption{Fractional contributions larger than $1\%$ to the total F production (modulated at 700~MV) as a function of rigidity. The top panel shows the ranking for CR element progenitors, while the bottom panel shows more details via the ranking of the 1-step and 2-step channels (reaction paths linking one isotopic progenitor to a CR isotope); `>2'-step channels (not shown) contribute to a total of $\sim10\%$, with about half of this number originating from the multi-step fragmentation of the Fe isotopes. In both plots, contributions starting from the same element share the same colours and line styles.
  \label{fig:xs_prodchannels}}
\end{figure}
We show in Fig.~\ref{fig:xs_prodchannels} the main F progenitors, from two slightly different perspectives. The top panel shows how much each CR element contributes to the total TOA flux of F--- the contributions are summed over ISM targets and account for the production of ghosts nuclei via the cumulative cross sections. The production from Ne is dominant ($\sim 30-40\%$), followed by Mg and Si ($\sim 20\%$), and then Fe ($\sim 10\%$). The next elements are S and Al, contributing at the percent level, and we then have a few sub-percent contributors (Na, Ca, Ar, P, Mn, Cl, and Cr in decreasing order of importance) whose total contribution reaches $\sim2\%$.
The low-rigidity decrease seen in the different contributions reflects the properties of the associated elemental fluxes: ratios of heavier-to-lighter primary fluxes decrease with decreasing rigidity \citep{2011A&A...526A.101P,2021PhRvL.126d1104A}, related to the growing destruction of CRs with their mass (the heavier the species, the larger its inelastic cross section). This pattern is seen in the various progenitors, being more or less marked according to the mass ordering of the elements (i.e. Ne, Mg, Al, Si, S, Fe).

We also show in the bottom panel of Fig.~\ref{fig:xs_prodchannels} a more detailed view of the isotopes of interest. Formally, for any given CR progenitor, one can list all the (possibly multi-step) reactions that link this progenitor to $^{19}$F. We can then rank the most important reactions \citep{2018PhRvC..98c4611G}. We dubbed these lists of reactions `channels', as they are not associated to a unique cross-section reaction (a ranking of the individual cross sections is also possible, but not shown here). In practice, we rank the 1-step and 2-step channels only. The most important progenitors from direct channels are $^{20}$Ne ($\sim 25\%$), $^{24}$Mg and $^{28}$Si (both at $\sim 20\%$), then $^{22}$Ne and $^{56}$Fe (both at $\sim7\%$), and other ones are at the percent level ($^{25,26}$Mg and $^{32}$S) or below (not shown). This ranking motivated the choice of the progenitors in App.~\ref{app:rescaling}, for the update of the most important reactions.
Beside the direct channels, two-step channels only reach the percent level, with a peak at a few GV\footnote{As underlined in \citet{2018PhRvC..98c4611G}, compared to 1-step production, the rigidity dependence of $n$-step channels follows $(1/K)^{n-1}\propto R^{-\delta(n-1)}$, with $\delta$ the diffusion slope.}. Although no multi-step fragmentation of $^{56}$Fe reaches the percent level individually, their multitude combine in a larger total contribution (also peaking at a few GV): the dashed-dotted magenta curve at 3~GV moves from $\sim1\%$ (direct contribution, bottom panel) to $\sim10\%$ (direct plus all 2-step contributions, top panel); the same trend, though with a more moderate impact of multi-step contributions, can be seen in $^{24}$Mg (long-dashed orange lines) and $^{28}$Si (dash-dotted magenta lines).

To conclude this section, at 10~GV, we find that 1-step channels contribute to $\sim70\%$ of the total F production, 2-step channels to $\sim20\%$, and `$>$2-step' channels to $\sim10\%$ of the total (and we recall that  multi-step contributions peak at a few GV). We can also tie $\sim62\%$ of the F production to 5 direct channels, namely from $^{20,22}$Ne, $^{24}$Mg, $^{28}$Si, and $^{56}$Fe), while $\sim 25\%$ originates from a few percent-level channels, and $\sim 13\%$ from hundreds of sub-percent level channels. We checked that these numbers only marginally depend on the production cross-section set considered (i.e. \optxii{}, \optxiiupxxii{}, and \optxxii{}).


\section{Primary fluxes}
\label{app:prim}

\begin{figure}[t]
\includegraphics[width=\columnwidth]{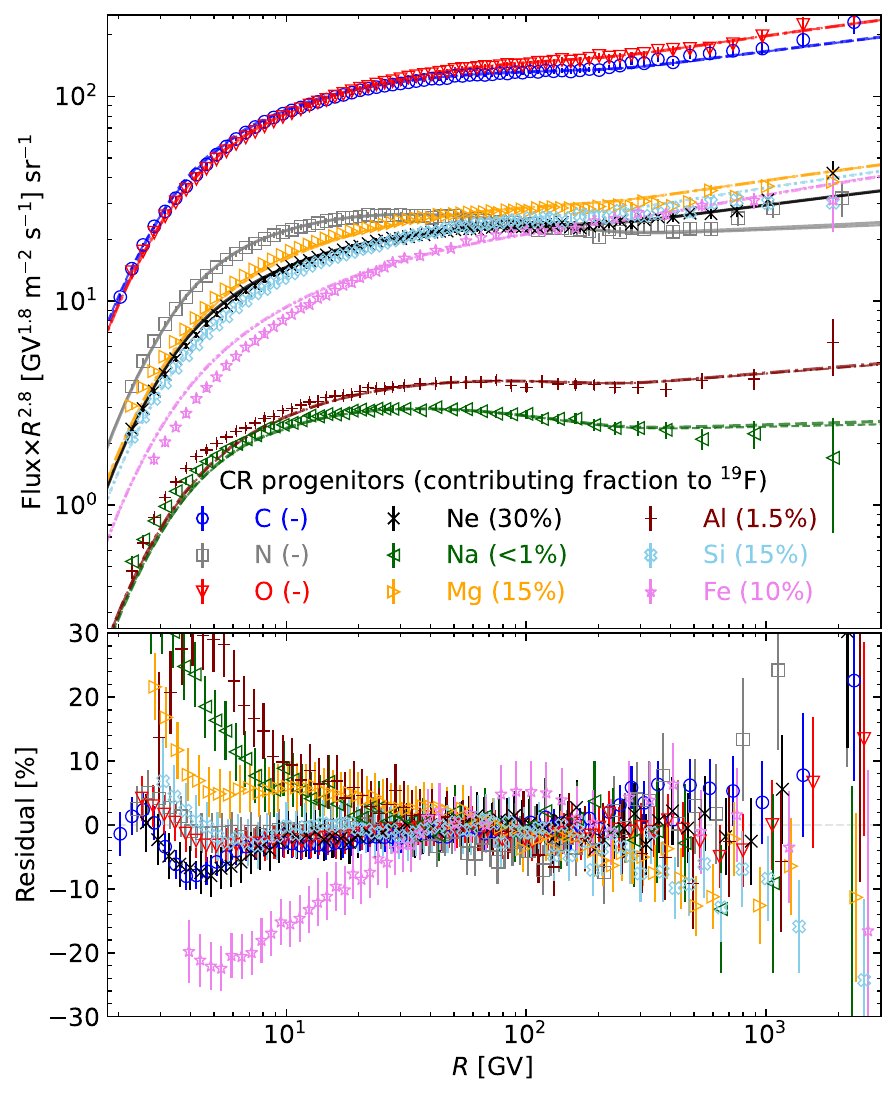}
  \caption{Comparison (top panel) and residuals (bottom panel) between the model calculation (lines) and the AMS-02 data \citep{2017PhRvL.119y1101A,2020PhRvL.124u1102A,2021PhRvL.126d1104A,2021PhRvL.127b1101A}, for the main progenitors of Li, Be, B, and F; we recall that this is not a fit to the data (see text). The numbers in parenthesis in the legend indicate the fractional contribution of these progenitors to the F production (as read from top panel of Fig.~\ref{fig:xs_prodchannels}).
   \label{fig:F_primfluxes_PurePL}}
\end{figure}

As highlighted in the main text, we do not fit the progenitors of Li, Be, B, and F in our runs. Instead, we adjust the individual elemental source abundance (and Ne isotopic source abundance) in order to match the data at a given rigidity. For AMS-02 data, this rigidity is taken to be the closest data point to 50~GV. As was shown in \citet{2019PhRvD..99l3028G}, this procedure gives an excellent match to the C, N, and O data. The same conclusions are drawn here in Fig.~\ref{fig:F_primfluxes_PurePL} for CNO (blue, red, and grey lines/symbols), although a few other elements show some discrepancies. At low rigidity, the model overshoots the Na and Al data (and to a lesser extent Mg) and undershoots Fe data, at the level of $\lesssim 30\%$; the Fe undershoot has been observed and discussed in \citet{2021PhRvD.103l3010S} and \citet{2021ApJ...913....5B}. It is beyond the scope of this paper to quantify this difference (statistically speaking) and to then discuss the possible origins of these mismatches.  We stress that a complete analysis should account for the covariance matrix of uncertainties in the data, improving the statistical agreement between the model and the data.

In the context of this analysis, it is enough to focus on the main progenitors of F, that is Ne, Mg, Si, and Fe. Their contributing fractions to the production of fluorine are reported in parenthesis in the legend of Fig.~\ref{fig:F_primfluxes_PurePL} (as taken form the top panel of Fig.~\ref{fig:xs_prodchannels}). We see that the main progenitors, Ne ($30\%$ contribution, black $\times$ symbol) and Si ($15\%$ contribution, empty cyan square), match very well the data. On the other hand, Mg ($15\%$, orange right-oriented empty triangle) and Fe ($10\%$, empty violet star) are off by $\lesssim20\%$. This makes their individual mismatch on F at the level of $\lesssim 30\%\times20\%\lesssim6\%$, but these two contributions cancel out because of their opposite signs. The remaining contributions from Na and Al ($1.5\%$) are too small to impact F, even though the model is $\lesssim 30\%$ below these data.
\begin{figure}[t]
\includegraphics[width=\columnwidth]{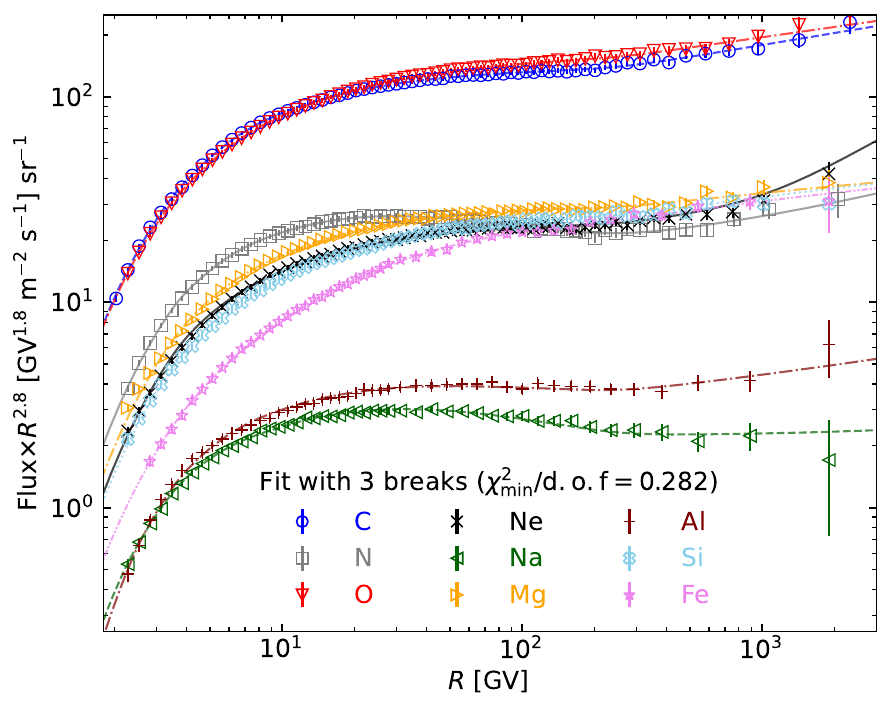}
  \caption{Same as Fig.~\ref{fig:F_primfluxes_PurePL}, but now fitting the primary fluxes assuming three breaks in their source spectra (see text); we do not show the residuals as they all overlap around zero. The model overfits the data with $\chimindof{}=0.282$.
  \label{fig:F_primfluxes_BrokenPL}
  }
\end{figure}

To further check this cancellation, we perform a dedicated fit of the primary source spectrum of C, N, O, Ne, Na, Al, Mg, Si, and Fe on the corresponding AMS-02 data. Instead of assuming a single power law (PL hereafter) as in the analysis presented in the main text, we follow the approach of \citet{2022ApJ...925..108B} and assume a broken power law (BPL hereafter) with three breaks:
\begin{equation}
Q \propto R^{-\alpha}\times \prod_{k=0}^2 \left(1+\left(\frac{R}{R_k}\right)^{\Delta_k/\eta_k}\right)^{\eta_k}.
\end{equation}
The smoothness parameters are fixed to $\eta_0=\eta_2=0.18$ (with positive values for $\Delta_0$ and $\Delta_2$ in the fit) and $\eta_1=-0.18$ (with negative values for $\Delta_1$ in the fit). With more than eight parameters per element---one normalisation per isotope, one slope $\alpha$ and three breaks $\Delta_k$ at three rigidities $R_k$---, it is no surprise that the model overfits the data ($\chimindof{}=0.282$). The fit and the data for the BPL configuration are shown in Fig.~\ref{fig:F_primfluxes_BrokenPL}; we do not report the values of the best-fit parameters (obtained assuming total uncertainties in quadrature on the data) as they are not relevant for the discussion.
We also stress that it is not clear whether the BPL configuration provides more trustworthy predictions, especially at high rigidity where the last break is dominated by a few data points (dominated by large statistical uncertainties). A dedicated analysis to assess the statistical relevance of the various breaks (in BPL) against some degree or full universality (in PL) in the primary spectra is necessary, but it goes beyond the scope of the paper. Importantly, such an analysis would have to include the covariance matrices of uncertainties on the data.

\begin{figure}[t]
\includegraphics[width=\columnwidth]{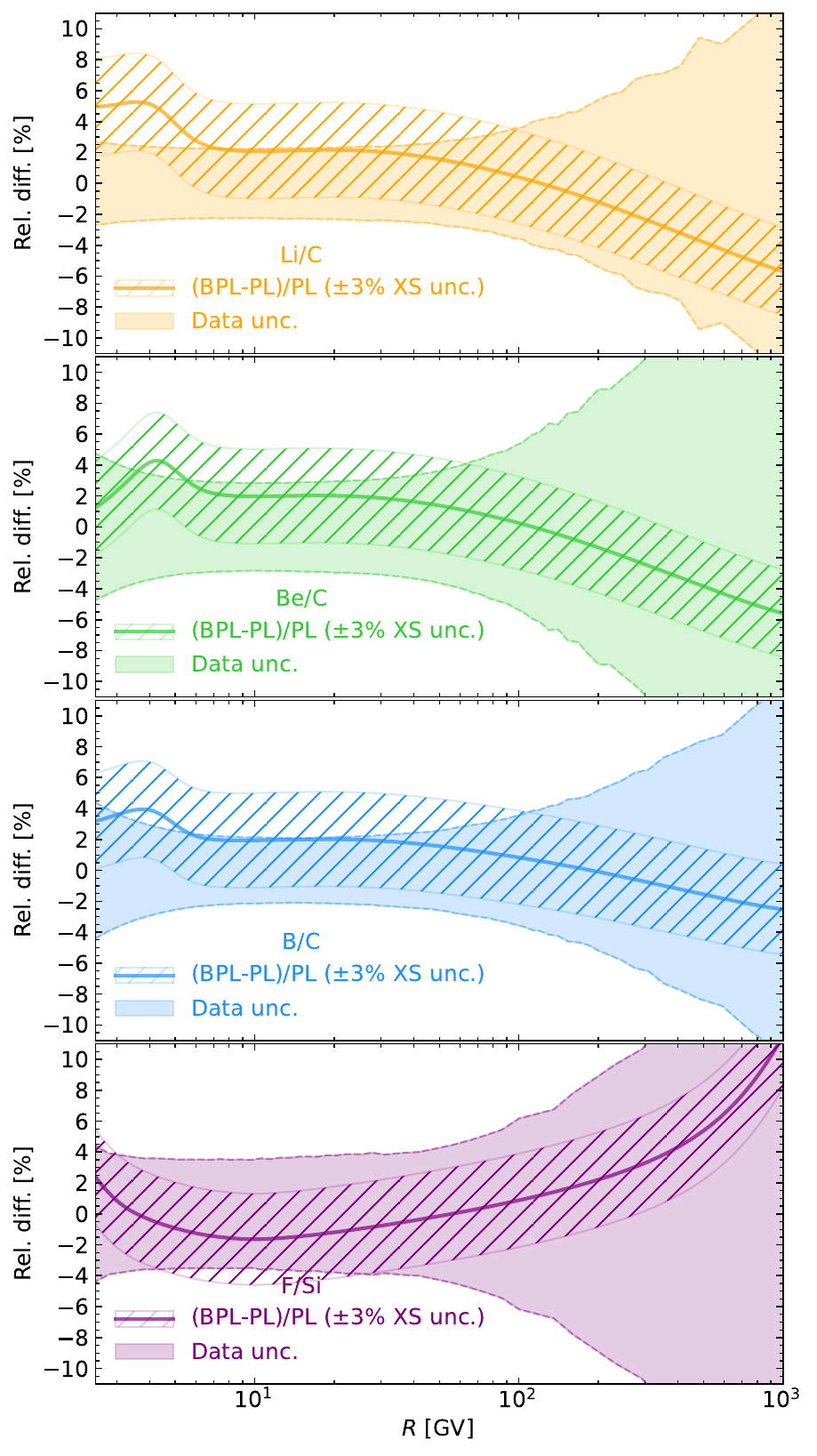}
  \caption{Relative difference between secondary-to-primary ratios calculated from the BPL and PL primary spectra, in solid lines (from top to bottom, Li/C, Be/C, B/C, and F/Si). The hatched area shows a $3\%$ shift related change in the production cross sections of the secondary in the numerator of these ratios. The shaded band shows the AMS-02 data relative  uncertainties (systematic and statistical uncertainties added in quadrature). We see that both areas overlap, showing that the difference between assuming PL or BPL does not impact our conclusions. We give more details in the discussion in the expanded text in App. A1.
   \label{fig:impact_fitprim}}
\end{figure}
We can now compare the secondary-to-primary ratios (of interest for this study) obtained with the BPL and PL assumptions in Fig.~\ref{fig:impact_fitprim}.
\begin{itemize}
  \item High-rigidity: above $\sim 100$~GV, the difference between the two calculations (solid lines) grows. In this regime, the difference (on the secondary-to-primary ratio calculation) can be interpreted as the propagation of the primary flux data uncertainties.\footnote{Indeed, with too many parameters on the model side and large statistical uncertainties on the data side, the BPL breaks (different breaks per elements) differ from the PL ones (universal break): for instance, the difference in the Ne flux (black line in Fig.~\ref{fig:F_primfluxes_BrokenPL} and \ref{fig:F_primfluxes_PurePL} respectively), the main progenitor of F, is responsible for the growing difference in the F/Si calculation.} In any case, the solid black line curve is within the secondary-to-primary AMS-02 data uncertainties (shaded band).

  \item Low-rigidity: below a few tens of GV, the difference between the BPL and PL calculation (solid lines) is larger than the data uncertainties (shaded area). However, the discrepancy is reconciled by a minimal change of the secondary production, shifting the solid lines by $\pm3\%$, as shown by the hatched band. This number is (i) of the order of the uncertainties obtained on the production cross section nuisance parameters (size of the ellipse in Fig.~\ref{fig:ellipses_SLIM}); (ii) of the order or smaller than the bias on these nuisance parameters (y-axis values of the centre of these ellipses in Fig.~\ref{fig:ellipses_SLIM}); (iii) for the Li case, much smaller than the dispersion related to the choice of the production cross-section set for the Li case (dispersion between the three orange ellipses in Fig.~\ref{fig:ellipses_SLIM}). This conservative hatched area and shaded one overlap, showing statistical consistency between the PBL and PL calculations.
\end{itemize}
For all these reasons, we conclude that while the BPL and PL approaches lead to few percent differences on the secondary-to-primary ratios (used in our analysis), these differences are either encompassed in the data uncertainties or  absorbed by a small shift of the production cross sections. This shift would certainly change the exact numbers provided in the main paper, but would not change any of our conclusions.

\end{appendix}
 
\end{document}